\documentclass[twocolumn, trackchanges]{aastex631}
\shorttitle{decomposing the spectrum of NGC 300 ULX-1}
\shortauthors{Kobayashi et al.}
\graphicspath{{./}}

\begin{document}

\title{Decomposing the Spectrum of Ultra-Luminous X-ray Pulsar NGC 300 ULX-1}

\author[0000-0001-7773-9266]{Shogo B. Kobayashi}
\email{shogo.kobayashi@rs.tus.ac.jp}
\affiliation{Department of Physics, Tokyo University of Science, 1-3 Kagurazaka, Shinjuku-ku, Tokyo 162-8601, Japan}

\author[0000-0001-6020-517X]{Hirofumi Noda}
\affiliation{Department of Earth and Space Science, Graduate School of Science, Osaka University, 1-1 Machikaneyama-cho, Toyonaka-shi, Osaka
560-0043, Japan}

\author[0000-0003-1244-3100]{Teruaki Enoto}
\affiliation{Department of Astronomy, Faculty of Science, Kyoto University, Kitashirakawa Oiwake-cho, Sakyo-ku, Kyoto 606-8502, Japan}

\author[0000-0001-8527-0496]{Tomohisa Kawashima}
\affiliation{Institute for Cosmic Ray Research, The University of Tokyo, 5-1-5 Kashiwanoha, Kashiwa, Chiba 277-8582, Japan}

\author[0000-0002-0700-2223]{Akihiro Inoue}
\affiliation{Center for Computational Sciences, University of Tsukuba, 1-1-1 Tennodai, Tsukuba, Ibaraki 305-8577, Japan}

\author[0000-0002-2309-3639]{Ken Ohsuga}
\affiliation{Center for Computational Sciences, University of Tsukuba, 1-1-1 Tennodai, Tsukuba, Ibaraki 305-8577, Japan}



\begin{abstract}
A phase-resolved analysis on the X-ray spectrum of Ultra-Luminous X-ray Pulsar (ULXP) NGC~300~ULX-1 is performed with data taken with XMM-Newton and NuSTAR on 2016 December 16th. In addition to the classical phase-restricting analysis, a method developed in active galactic nuclei studies is newly employed for ULXP. It has revealed that the pulsation cycle of the source can be divided into two intervals in terms of X-ray variability. This suggests the rotating flow consists of at least two representative emission regions. Furthermore, the new method successfully decomposed the spectrum into an independent pair in each interval. One is an unchanging-component spectrum that can be reproduced by a standard disk model with a $720^{+220}_{-120}$~km inner radius and a $0.25\pm0.03$~keV peak temperature. The other is the spectrum of the component that coincides with the pulsation. This was explained with a Comptonization of a $0.22^{+0.2}_{-0.1}$~keV blackbody and exhibited a harder photon index in the brighter phase interval of two. The results are consistent with a picture that the pulsating emission originates from a funnel-like flow formed within the magnetosphere, and the inner flow exhibiting a harder continuum is observed exclusively when the opening cone points to the observer.

\end{abstract}



\section{Introduction} \label{sec:intro}
Ultra-Luminous X-ray sources (ULXs) are X-ray bright point sources frequently found at the off-nucleus regions of galaxies with high star formation rates, such as the (interacting) spirals, dwarfs, and starbursts (\citealt{Kaaret2017, Walton2021, King2023}). Since their X-ray luminosity, $10^{39.5\--41}$~erg~sec$^{-1}$, well exceeds the Eddington limit of stellar-mass ($10M_{\rm \odot}$) black holes, they are often regarded as possible candidates for intermediate-mass ($10^{2\--3}M_{\rm \odot}$) black holes (e.g., \citealt{Makishima2000}) or stellar-mass objects accreting matters at well above their Eddington rate (e.g., \citealt{Mineshige2007}). 

Although the true nature of ULXs has been under discussion since its discovery in the 1980s \citep{Fabbiano1987}, the epoch-making detection of $1.4$~sec X-ray pulsation from M82~X-2 \citep{Bachetti2014} has revealed that at least some fractions of ULXs do harbor neutron stars as their central object accreting at $\sim 100$ times the Eddington limit. At present, 9 (8 extra-Galactic and 1 Galactic) ULXs are confirmed to be containing neutron stars as their accretor \citep{Bachetti2014, Furst2016, Israel2017a, Israel2017b, Tsygankov2017, Carpano2018, Wilson_Hodge2018, Sathyaprakash2019, Castillo2020, Chandra2020, Vasilopoulos2020}. Since the ULX Pulsars (ULXPs) are those of limited systems that are firmly regarded as accreting at well above their Eddington rate, these are intensively studied as ideal systems to unravel the poorly understood nature of the super-critical accretion flows. 

X-ray spectral analysis often plays a significant role in studying accretion physics and geometry of mass-accreting objects. However, those in ULXs, including ULXPs, generally provide few clues. This is mainly because ULXPs tend to exhibit continuum-dominated spectra with few characteristic features, allowing multiple spectral models to explain the same spectrum with nearly identical statistics. Therefore, it is crucial in ULXPs studies to somehow grasp the actual spectral shapes of the components forming the continuum, and an analysis method ideal for such an objective is available from studies on Active Galactic Nuclei (AGN).

Following \citet{Churazov2001} and \citet{Taylor2003}, \citet{Noda2014} (see also \citealt{Noda2011, Noda2013a}) introduced a method that extracts the spectral shape of components that forms the original X-ray spectrum of the source. It relies on the correlation between the count rates of two energy bands, one from a fixed energy band and the other from an arbitrary part of the rest. By evaluating how these two correlates as the X-ray intensity of the source varies, one can directly derive the exact count rate contribution of two consisting components at that energy band; one that changes its intensity in coincidence with the X-ray variability and the other that does not. Repeating this procedure in various energy bands, the authors successfully decomposed a featureless spectrum of an AGN into two additional ones without relying on any physical models. Since these two newly-extracted spectra are additionally available for the model fittings, the method provided the authors with more stringent restrictions to their spectral modeling, and the same asset can be expected in the ULX studies. Especially in rotating neutron stars like ULXPs, their X-ray pulse periods, which are unavailable in black holes like those in AGNs, enable us to distinguish emission components originating from pulsating regions bound to the dipole magnetic field of a neutron star and those from the unbound one. Our objective in this work is to apply this method introduced by \citet{Noda2014} to the pulsation of an ULXP for the first time and untangle the model degeneracy that has been present in ULX studies for nearly a decade.

NGC~300~ULX-1 (hereafter, ULX-1) is a ULXP residing in a nearby (1.9~Mpc; \citealt{Gieren2005}) spiral galaxy NGC~300. Its X-ray emission was first detected with Chandra as a weak ($\sim6\times10^{35}$~erg~sec$^{-1}$; \citealt{Binder2011}) source associated with a super-nova imposter SN2010da \citep{Monard2010} and suddenly became bright as a ULX ($>10^{39}$~erg~sec$^{-1}$) in 2016. The object turned out to be a X-ray pulsar rotating with a spin period of $\sim 31.7$~sec and also spinning up with a significant rate of $5.6\times10^{-7}$~sec~sec$^{-1}$ \citep{Carpano2018}. Despite the highest spin-up rate among ULXPs, ULX-1 is still the one that rotates at the longest period ($\sim17$~sec in the latest observation in 2018; \citealt{Vasilopoulos2019}). The method mentioned above gives higher resolution if one can divide the spin period into a large number of sections with each containing a sufficient photon statistic. Hence, its relatively long spin period among ULXPs makes the source ideal for our study

\section{Observation and Data Reduction} \label{sec:observation}
In general, phase-resolved spectral analysis of neutron star X-ray binaries requires high photon statistics, time resolution, and wide energy band coverage. Hence, we utilize the large effective area and short read-out time of the XMM-Newton \citep{Jansen2001} European Photon Imaging Camera (EPIC) and the high-energy capability of the NuSTAR \citep{Harrison2013} Focal Plane Module (FPM). In the present study, we revisit the data sets utilized in several works (e.g., \citealt{Carpano2018, Walton2018b, Koliopanos2019}). They were taken with XMM-Newton and NuSTAR simultaneously from 2016 December 16th with the longest total duration among the ones currently available ($\sim320$~ks).

Throughout the observation, three EPIC instruments MOS1, MOS2, and pn were operating normally in full-window mode. Since this study requires a large effective area, we utilized only pn in the present analysis. All data screening processes are carried out with software included in Science Analysis System version 19.0.0 and the current calibration file updated on 2021/12/3. The pipeline processes such as gain calibration and removal of bad quality events are done by \texttt{epchain} with default criteria established by the XMM-Newton instrumental team.
The spectrum and light curve of ULX-1 are extracted from an on-source circular region with a $30''$ radius, while those of backgrounds are from a $60''$ radius circle placed $\sim 2.5'$ off from ULX-1 wherein no apparent X-ray sources are detected.

Two X-ray detectors on board NuSTAR, FPM-A and FPM-B, were also operating normally throughout the observation. All screening processes were carried out by using software included in High Energy Astronomy SOFT version 6.28 and the calibration data base updated on 2022/01/06. Basic pipeline processes such as bad-grade event reduction, mast and spacecraft attitude correction, and discarding events within South Atlantic Anomaly are done via \texttt{nupipeline} command. We set all of the parameters in this procedure to the recommended values set by the NuSTAR instrumental team. Secondary products generated from the pipeline-processed event data (such as the X-ray spectra, light curves, and response matrix files) are generated with the \texttt{nuproducts} command. The source spectra and light curves are extracted from a $30''$ radius circular region, and those of backgrounds are from a $60''$ radius one with a $\sim3'$ offset from ULX-1.

The arrival time of each X-ray event in both EPIC-pn and FPM data is corrected for the barycentric coordinate of the solar system with the JPL planetary ephemeris DE-200 \citep{Standish1990}. Here, we adopted a source coordinate of $({\rm RA},~{\rm DEC})=(3.77^{\circ},~ -37.70^{\circ})$, which is the direction toward ULX-1. According to \citet{Walton2018b}, the measured initial pulsation period at 57738.65732~MJD and its derivative are $P=31.7183411308$~sec and $\dot{P}=-5.563\times10^{-7}$~sec/sec, respectively. We used this result to calculate the pulsation phase of each event in the data set. 

\section{Analysis and Results} \label{sec:results}
\subsection{Light Curve and Pulse Profile} 
\begin{figure}
    \centering
    \includegraphics[width=\columnwidth]{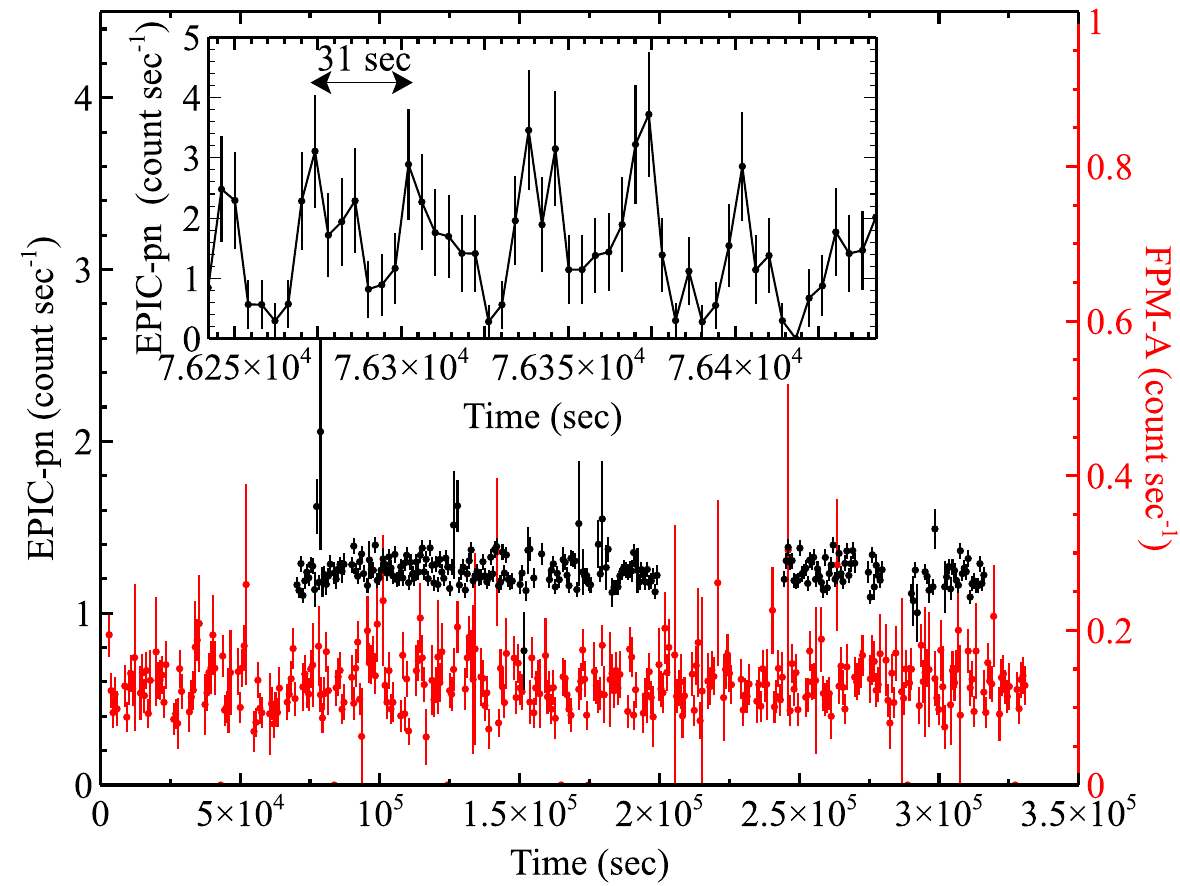}
    \caption{Background subtracted light curves of ULX-1 taken with EPIC-pn ($0.3\--10.0$ keV: black) and FPM-A ($3\--25$~keV: red). The zero point of the horizontal axis corresponds to the beginning of the NuSTAR observation (MJD 57738.6573). Those in the larger panel has a full length of the observation with 700~sec width per bin, whereas one in the other panel is a 200~sec fraction of particular interval in the EPIC-pn observation with a 4 sec bin width. }
    \label{fig:light_curve}
\end{figure}
In Figure \ref{fig:light_curve}, we present background subtracted-light curves of ULX-1. Those in the larger panel have a length of the overall observation with a bin width of 700~sec, and one in the top left is a small portion (200~sec) of XMM-Newton pn observation with a bin width of 4~sec. We can confirm the presence of coherent variability with the $\sim31$-sec pulsation period in the 4~sec bin light curve, whereas nothing significant can be found in the others with the longer-time bins. Thus, the data set contains no apparent X-ray variability but from pulsation, making these data sets ideal to study the pure variability of the pulsating component.

\begin{figure*}
\begin{minipage}[b]{0.5\linewidth}
     \centering
    \includegraphics[width=\columnwidth]{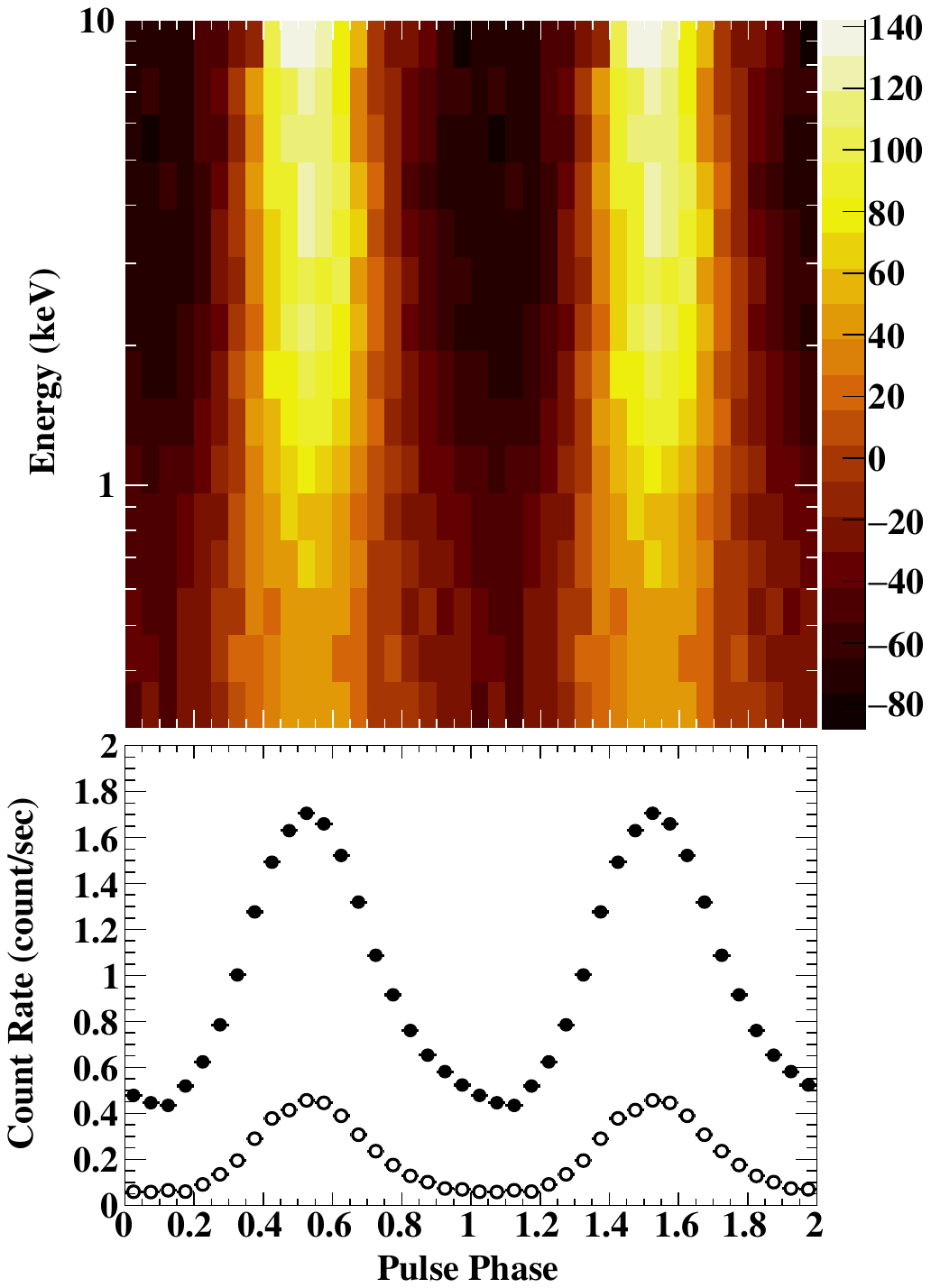}
\end{minipage}
\begin{minipage}[b]{0.5\linewidth}
    \centering
    \includegraphics[width=\columnwidth]{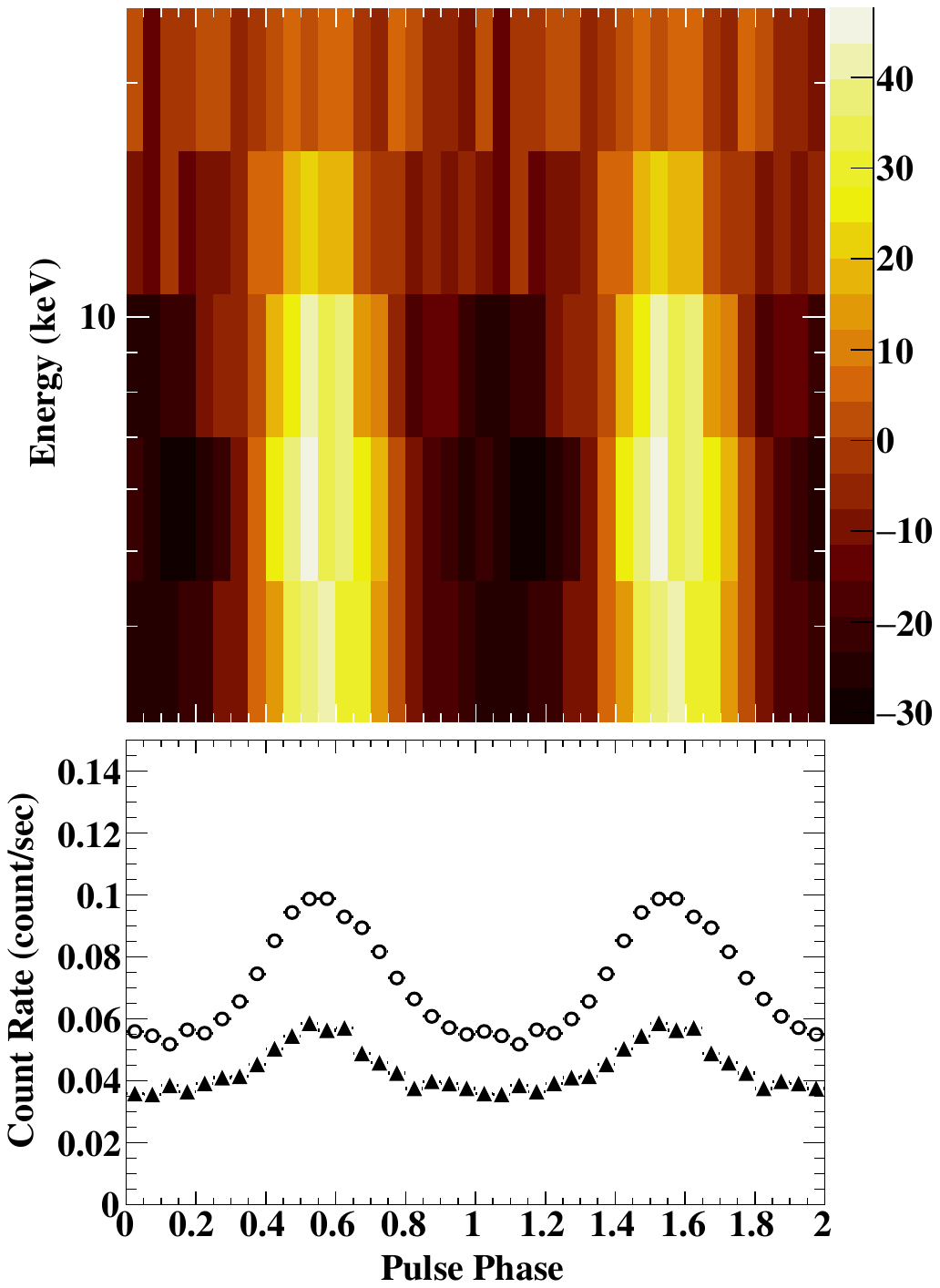}
\end{minipage}
    \caption{X-ray pulse profile (bottom) and its energy dependency (top) obtained from the $0.3\--10$~keV band of XMM-Newton EPIC-pn (left) and the $3\--25$~keV of NuSTAR FPM-A (right). For clarity, only FPM-A is presented for the NuSTAR data. The color scales of the top panels represent the fluctuation over the average count rate in percent. The count rates in the bottom panels are the average over respective energy bands; $0.3\--3$~keV (filled circles), $3\--10$~keV (open circles), and $10\--25$~keV (filled triangles). The data points of $10\--25$~keV are scaled by a factor of four for clarity}.
    \label{fig:pulse_profile}
\end{figure*}

 Figure \ref{fig:pulse_profile} represents how the X-ray intensity of NGC~300~ULX-1 varies in terms of the pulse phase and energy. As described in section \ref{sec:intro}, we folded the entire data set with the initial period and spin-up rate derived by \citet{Walton2018b}, which are 31.7183411308~sec (reference epoch is 57738.65732~MJD) and $-5.563\times10^{-7}$~sec~sec$^{-1}$, respectively. The pulse profile is relatively sinusoidal, which is consistent with the previous results (e.g., \citealt{Carpano2018}), and peaked at $\phi=0.5\--0.6$ throughout $0.3\--25$~keV, suggesting a ``single zone'' hot spot.

\begin{figure}
    \centering
    \includegraphics[width=\columnwidth]{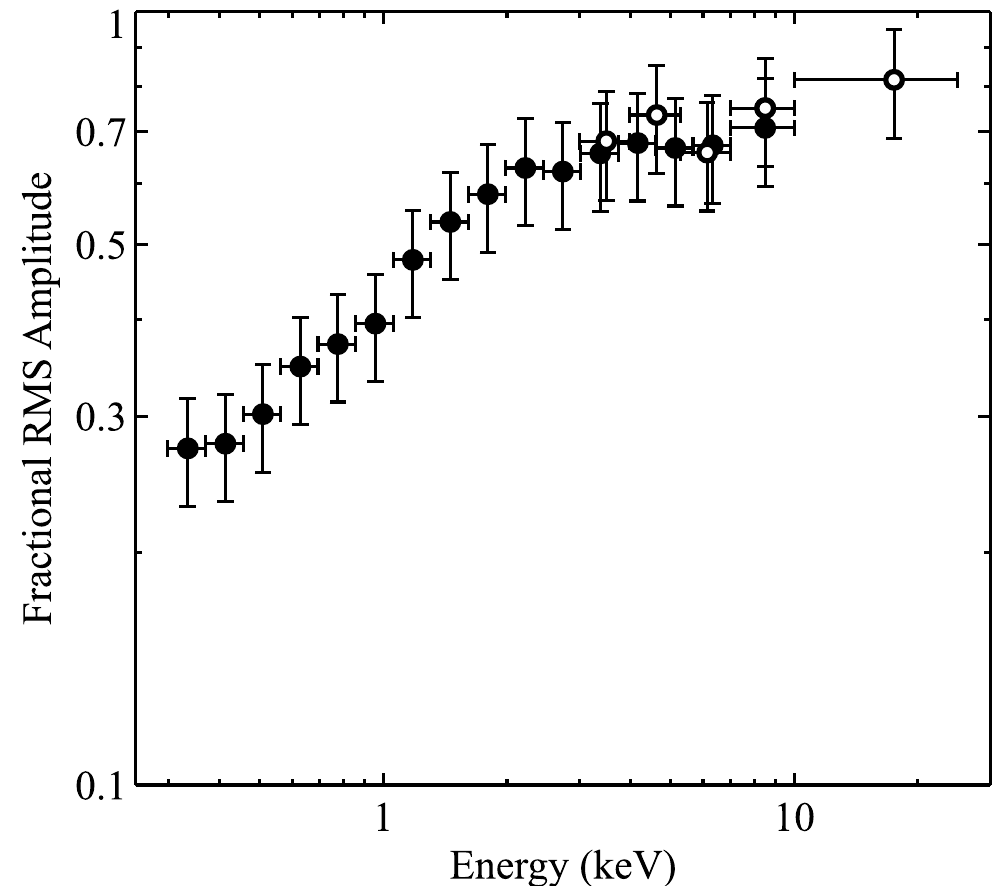}
    \caption{Energy dependency of the fractional RMS amplitude over the pulsation cycle of ULX-1 calculated from the XMM-Newton PN detector (filled circles) and FPM (open circles) data. For clarity, only the data from FPM-A are shown for NuSTAR.}
    \label{fig:pulse_fraction}
\end{figure}

Figure \ref{fig:pulse_fraction} presents an energy dependency of the fractional Root Mean Square (RMS) amplitude, $\sqrt{S^{2}/\bar{x}}$, where $S^{2}$ and $\bar{x}$ are the variance and average of the count rate over the pulsation cycle, respectively. The fractional RMS amplitude becomes larger as the energy increases, reaching $\ge 70\%$ above $3$~keV. This is consistent with the previous results that used the same data sets as the present study (e.g., \citealt{Carpano2018, Vasilopoulos2019}). Hence, the spectrum of NGC~300~ULX-1 is expected to be dominated by the pulsating emission component at a higher energy band.

\subsection{Count-Count Correlation with Positive Offset (C3PO) Method \label{sec:c3po}}
Just like those in ULXPs, AGNs in Seyfert 1 galaxies also tend to exhibit featureless X-ray spectra, allowing multiple physical models to reproduce them without any significant statistical differences. Therefore, adding restrictions to models has been crucial, and spectral variability is one of the helpful tools to resolve this model degeneracy.

\citet{Noda2014} introduced a method called Count-Count Correlation with Positive Offset (C3PO), which utilizes the characteristic X-ray variability of AGN to extract a pair of spectra that composes the original spectrum without relying on any physical models. One is from the component that accounts for the variability, and the other is from the unchanging one. The method is based on a correlation between the count rates of two energy bands. One is from a fixed energy range defined as a reference band, and the other is from a test band that is an arbitrary part of the rest. If we plot the test-band count rate against that of the reference band, the data points ought to exhibit a certain locus whose shape depends on how the spectral shape changes in time. The simplest example is when the variable component changes only its intensity. In this case, the data points will form a straight line on the count rate v.s. count rate plot (CCP) plane, and a product of its slope and the count rate in abscissa represents the actual count rate of the variable component at that energy band. Furthermore, if the locus shows any positive offset at the zero point of the reference band, then the source spectrum is likely to contain a component that does not correlate to the variability with a count rate equivalent to that very offset value.

\citet{Noda2014} found that the CCPs at the bright state of a highly variable AGN, NGC 3227, form straight loci, and each of them shows an apparent positive offset if they extrapolate the data down to the count rate where that of the reference band reaches zero. It clearly suggests, as described above, that the X-ray spectral continuum of NGC 3227 consists of at least two components. One is something uncorrelated (or stable) against the variability representing the positive offset in CCP, and the other is the variable component that changes only its intensity in time. The authors fitted each CCP with a linear function expressed as
\begin{equation}
    y=ax+b,
    \label{eq:linear}
\end{equation}
where $x$, $y$, $a$, and $b$ are the reference-band count rate, test-band count rate, slope, and y-intercept value, respectively. Here, $a$ and $b$ are set as free parameters, and the derived value $ax_0$ ($x_0$ is the average of $x$ used in the fitting) and $b$ represent the actual count rate of the variable and stable component at that energy band, respectively. By repeating this procedure in the other energy bands, \citet{Noda2014} successfully extracted the spectrum of the stable component with a thermal emission shape and a variable one with an absorbed power-law figure hidden under the featureless X-ray spectrum of NGC 3227. In the present study, we apply the same C3PO method as \citet{Noda2014} to NGC~300~ULX-1, but in a pulse phase-dependent manner.
\begin{figure}
    \centering
    \includegraphics[width=\columnwidth]{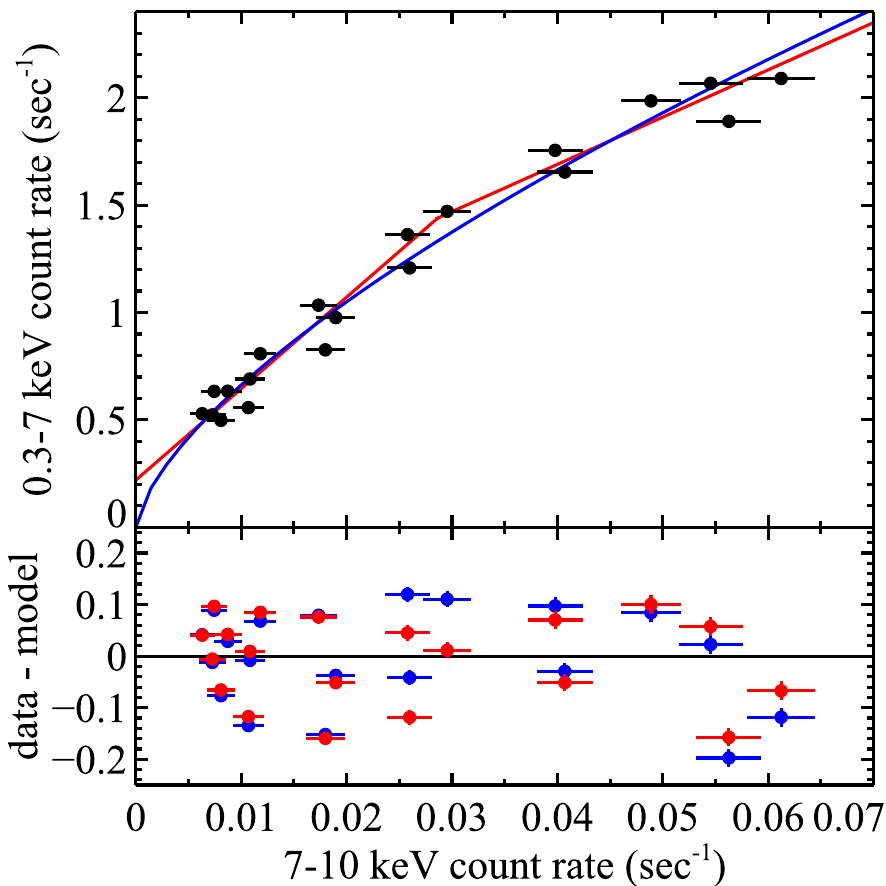}
    \caption{A CCP obtained from the $0.3\--7$~keV band of the XMM-Newton PN data (top) and its residuals from the respective model function (bottom). The red and blue solid lines are the best-fit breaking-linear function and exponential function, respectively. The colors of the residual data points are in correspondence with those of the function curves above. The data points with the highest and lowest count rate correspond to the peak and bottom of the pulse, respectively.}
    \label{fig:brek_vs_power}
\end{figure}

Figure \ref{fig:brek_vs_power} presents the extracted CCP of ULX-1. Taking the high photon statistics of XMM-Newton and the highest pulse fraction (or variability) at $> 7$~keV (Figure \ref{fig:pulse_fraction}) into account, we select $7.0\--10.0$~keV of XMM-Newton pn as the reference band (horizontal axis). To grasp the overall behavior of the CCP, we maximized the statistics by designating the rest of the entire energy band, $0.3\--7.0$~keV, as the test band (vertical axis). Instead of the raw count rates utilized in \citet{Noda2014}, each data point in this CCP represents an average count rate within a divided portion ($1/20=0.05$ cycle per data point in this case) of the $\sim31.7$~sec pulsation cycle. In short, the CCPs in the present paper reflect how the count rate in each energy band varies as a function of the pulsation phase.

The CCP of ULX-1 is forming a rather straight locus with a ``break'' at $\sim0.03$~count~sec$^{-1}$, which corresponds to a pulse phase of $\phi\sim0.4$ and $\phi\sim0.75$. It suggests that the variable component, emission bound to the pulsating accretion flow, increases its X-ray intensity without changing its spectral shape within phase intervals below or above this break. Since the data above the bent are forming a locus with a shallower slope, the appearance of the spectra should be different between these two phases.

A similar breaking CCP is also reported in NGC~3227 by \citet{Noda2014}. Instead of explaining the entire data points with a single linear line as \citet{Noda2011} and \citet{Noda2013a} did, the authors tested two alternative functions that could fit such curving CCPs. One is a piecewise-segmented linear function that consists of a pair of linear functions individually explaining the data points below and above the break separately. The other is a single power-law function expressed as $y=Mx^{N}$ (e.g., \citealt{Uttley2005}), where $M$ and $N$ are left free to vary. Following \citet{Noda2014}, we tested these two possible solutions as shown in Figure \ref{fig:brek_vs_power}. The piecewise-segmented linear function is defined around the breaking point $x_{\rm b}$ as,
\begin{eqnarray*}
    y = a_1x+b_1~~(x\le x_{\rm b})\\
    y = a_2x+b_2~~(x > x_{\rm b}) \\
    x_{\rm b} = (a_1 - a_2)/(b_2 - b_1),
\end{eqnarray*}
where $a_i$ and $b_i$ ($i=1,2$) are the slopes and intercepts of individual linear functions, respectively. Hence, the best-fit slopes and intercepts automatically derive $x_b$. In the regression analysis, we utilized the ROOT analysis package developed by CERN. The errors in $x$ are projected to the $y$-axis direction to take contributions from both $x$ and $y$ into account by calculating chi-square at each data point as $\chi^2 = (y-f(x))/(\sigma_y^2+\sigma_x^2f'(x)^2)$, where $\sigma_x$, $\sigma_y$, $f(x)$, and $f'(x)$ are the errors in $x$ and $y$, the fitting function, and the derivative of the function, respectively. 

Although the piecewise linear function gave a slightly better fit ($\chi^2/$degree of freedom $=29.2/16$) than the single power-law function ($\chi^2/$degree of freedom $=34.2/18$), the difference is rather marginal. If we calculate the Bayesian Information Criteria, which can be expressed using the likelihood function $L$, number of parameter $k$, and number of data points $n$ as $-2\ln L+k\ln n$, the difference between the models is $<1$. According to criteria by \citet{Kass1995}, this is insufficient to claim that the piecewise linear function is statistically preferable over the power-law function.

Unlike \citet{Noda2014}, we were unsuccessful in statistically ruling out the power-law function due to the limited number of data points in the CCP. Since the difference between the two functions becomes significant around the breaking point, it may be possible to distinguish one from another by observing ULX-1 with a long exposure or instruments with larger effective areas. We leave this as a future work and adopt the results on piecewise linear function as a working hypothesis of the following analysis. The best-fit value of the breaking count rate is $0.028$ count~sec$^{-1}$, and hence we hereafter refer to the phase below this count rate ($0.0 \le \phi \le 0.4$ and $0.75 \le \phi \le 1.0$) as ``faint phase'' and above it ($0.4 < \phi <0.5$) as ``bright phase''.

\begin{figure*}
    \centering
    \includegraphics[width=\textwidth]{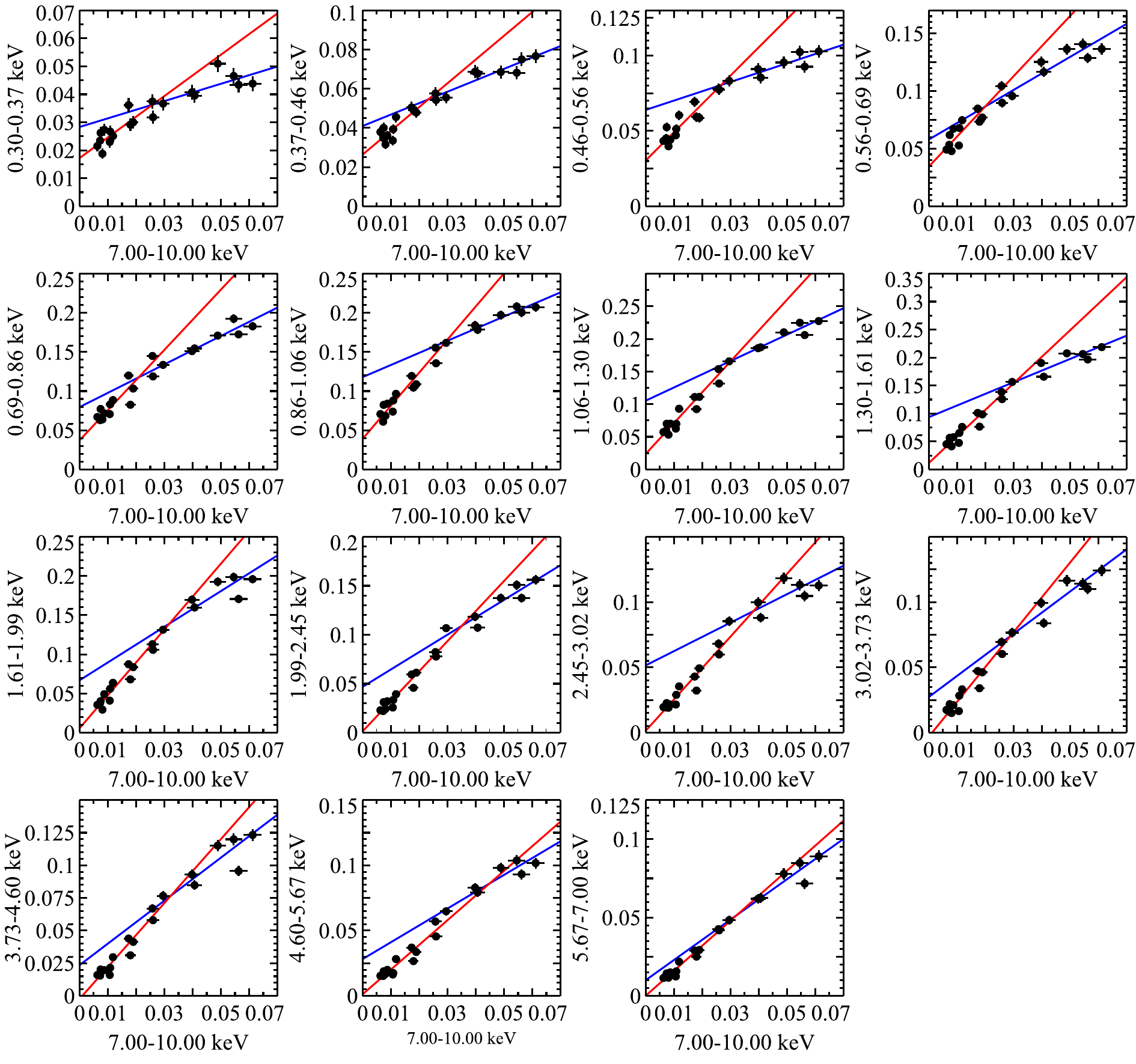}
    \caption{XMM-Newton pn phase-resolved Count-Count Plot of ULX-1. The red and blue solid lines represent the best-fit polynomial functions. The data points with the highest and lowest count rate correspond to the peak and bottom of the pulse, respectively.}
    \label{fig:ccplot}
\end{figure*}
\begin{figure*}
    \centering
    \includegraphics[width=\textwidth]{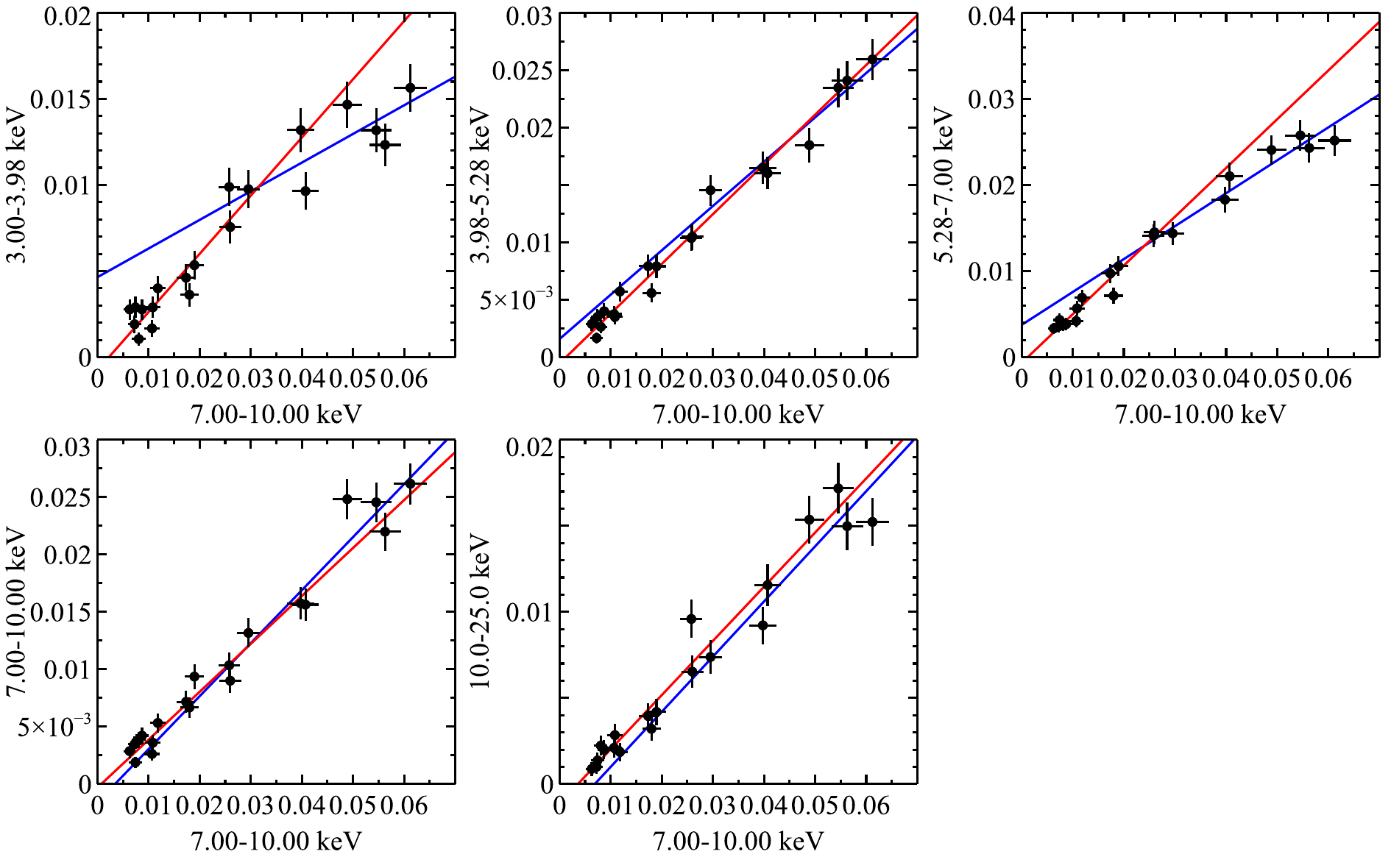}
    \caption{Same as Figure \ref{fig:ccplot}, but vertical axis represents the count rate of NuSTAR FPM-A}
    \label{fig:ccplot_fpma}
\end{figure*}
To extract the spectra of stable and variable components, we divided the $0.3\--7.0$~keV of XMM-Newton pn and $3.0\--25.0$~keV of NuSTAR FPM into 15 and 5 sub-bands, respectively, and created CCPs for each using $7.0\--10.0$~keV count rate of XMM-Newton pn as a reference (Figures \ref{fig:ccplot}, \ref{fig:ccplot_fpma} and \ref{fig:ccplot_fpmb}). The sub-bands are defined as those consisting of data points that contain at least 20 counts and have logarithmically-equal width within the individual instruments except for the $10\--25$~keV bands of NuSTAR FPMs. Thus, the same breaking feature as Figure \ref{fig:brek_vs_power} is present in the other energy band. The change in slope is apparent in the lower energy bands and becomes more ambiguous as the energy of the test band reaches 10~keV or higher.

 As \citet{Noda2014}, we fit the data points with count rates below and above the break in each CCP separately with the respective linear functions shown in blue and red solid lines in Figures \ref{fig:ccplot}, \ref{fig:ccplot_fpma}, and \ref{fig:ccplot_fpmb}. The goodness of fit and obtained parameters are summarized in Table \ref{tab:xmm_ccp_fit_result}. The pairs of linear functions successfully reproduced the individual CCPs, and some exhibited non-zero y-intercept values. Hence, following the procedure described above, we can generate two pairs of spectra from the best-fit values for each phase. Combining XMM-Newton and NuSTAR, we obtained a stable-component spectrum with 7 bins in $0.30\--1.61$~keV via data points in the faint phase area and another with 20 bins in $0.3\--10$~keV from those on the other side. As for the variable component, we obtained a spectrum with 25 bins between $0.3\--25$~keV from the faint phase, whereas those from the bright phase produced one with 22 valid bins ranging from $0.56$~keV to $25$~keV.

\begin{deluxetable*}{cccccccc}
\tablenum{1}
\tablecaption{Results on linear-function fit of CCPs \label{tab:xmm_ccp_fit_result}}
\tablewidth{0pt}
\tablehead{
& \multicolumn{3}{c}{below the break} & & \multicolumn{3}{c}{above the break}\\
 \cline{2-4} \cline{6-8}
\colhead{Energy band} & \colhead{slope} & \colhead{offset} & \colhead{reduced $\chi^2 (\nu)$} &  & \colhead{slope} & \colhead{offset} & \colhead{reduced $\chi^2 (\nu)$}}
\startdata
    \multicolumn{8}{c}{XMM-Newton EPIC-PN}\\
    $0.30\--0.37$~keV & $0.7\pm0.1$ & $0.017\pm0.002$ & 1.706 (11) & & $0.3\pm0.1$ & $0.028\pm0.005$ & 1.731 (5)\\
    $0.37\--0.46$~keV & $1.2\pm0.2$ & $0.026\pm0.002$ & 1.330 (11) & & $0.6\pm0.1$ & $0.041\pm0.006$ & 0.891 (5)\\
    $0.46\--0.56$~keV & $1.9\pm0.2$ & $0.030\pm0.003$ & 1.591 (11) & & $0.6\pm0.2$ & $0.064\pm0.008$ & 0.882 (5)\\
    $0.56\--0.69$~keV & $2.6\pm0.3$ & $0.034\pm0.003$ & 2.373 (11) & & $1.4\pm0.2$ & $0.06\pm0.01$ & 2.092 (5)\\
    $0.69\--0.86$~keV & $3.8\pm0.4$ & $0.037\pm0.005$ & 2.354 (11) & & $1.8\pm0.3$ & $0.08\pm0.01$ & 1.135 (5)\\
    $0.86\--1.06$~keV & $4.2\pm0.4$ & $0.039\pm0.005$ & 1.554 (11) & & $1.5\pm0.3$ & $0.12\pm0.01$ & 0.470 (5)\\
    $1.06\--1.30$~keV & $4.7\pm0.4$ & $0.024\pm0.005$ & 1.480 (11) & & $2.0\pm0.3$ & $0.11\pm0.01$ & 0.723 (5)\\
    $1.30\--1.61$~keV & $4.7\pm0.04$ & $0.011\pm0.005$ & 1.457 (11) & & $2.1\pm0.3$ & $0.09\pm0.01$ & 2.178 (5)\\
    $1.61\--1.99$~keV & $4.2\pm0.4$ & $0.006\pm0.004$ & 1.218 (11) & & $2.1\pm0.3$ & $0.09\pm0.01$ & 2.996 (5)\\
    $1.99\--2.45$~keV & $3.1\pm0.3$ & $0.001\pm0.003$ & 0.979 (11) & & $1.8\pm0.2$ & $0.05\pm0.01$ & 1.620 (5)\\
    $2.45\--3.02$~keV & $2.4\pm0.2$ & $0.001\pm0.003$ & 1.199 (11) & & $1.1\pm0.2$ & $0.05\pm0.01$ & 2.718 (5)\\
    $3.02\--3.73$~keV & $2.7\pm0.2$ & $-0.003\pm0.003$ & 1.382 (11) & & $1.6\pm0.2$ & $0.04\pm0.01$ & 1.759 (5)\\
    $3.73\--4.60$~keV & $2.4\pm0.2$ & $-0.002\pm0.003$ & 1.271 (11) & & $1.6\pm0.2$ & $0.02\pm0.01$ & 3.267 (5)\\
    $4.60\--5.67$~keV & $1.9\pm0.2$ & $0.001\pm0.002$ & 1.538 (11) & & $1.3\pm0.2$ & $0.03\pm0.01$ & 1.152 (5)\\
    $5.67\--7.00$~keV & $1.6\pm0.2$ & $4.9\times10^{-5}\pm0.002$ & 0.672 (11) & & $1.3\pm0.2$ & $0.010\pm0.008$ & 1.178 (5)\\
    \multicolumn{8}{c}{NuSTAR FPM-A}\\
    $3.0\--4.0$~keV & $0.34\pm0.04$ & $-0.0008\pm0.0005$ & 1.723 (11) & & $0.17\pm0.05$ & $0.005\pm0.002$ & 1.612 (5)\\
    $4.0\--5.3$~keV & $0.43\pm0.05$ & $-0.0005\pm0.0006$ & 1.612 (11) & & $0.39\pm0.07$ & $0.0016\pm0.003$ & 0.790 (5)\\
    $5.3\--7.0$~keV & $0.57\pm0.06$ & $-0.0007\pm0.0008$ & 0.534 (11) & & $0.38\pm0.07$ & $0.004\pm0.003$ & 0.622 (5)\\
    $7.0\--10.0$~keV & $0.42\pm0.05$ & $-0.0003\pm0.0006$ & 1.076 (11) & & $0.46\pm0.07$ & $-0.002\pm0.03$ & 1.198 (5)\\
    $10.0\--25.0$~keV & $0.31\pm0.04$ & $-0.0011\pm0.0005$ & 0.945 (11) & & $0.32\pm0.06$ & $-0.002\pm0.003$ & 1.095 (5)\\
    \multicolumn{8}{c}{NuSTAR FPM-B}\\
    $3.0\--4.0$~keV & $0.27\pm0.04$ & $0.0003\pm0.0005$ & 0.869 (11) & & $0.16\pm0.05$ & $0.005\pm0.002$ & 0.693 (5)\\
    $4.0\--5.3$~keV & $0.47\pm0.06$ & $-0.0005\pm0.0007$ & 1.134 (11) & & $0.30\pm0.07$ & $0.005\pm0.003$ & 2.620 (5)\\
    $5.3\--7.0$~keV & $0.38\pm0.06$ & $0.0018\pm0.0007$ & 1.238 (11) & & $0.39\pm0.07$ & $0.003\pm0.003$ & 0.765 (5) \\
    $7.0\--10.0$~keV & $0.48\pm0.06$ & $-0.0011\pm0.0007$ & 1.012 (11) & & $0.40\pm0.07$ & $0.0004\pm0.003$ & 1.324 (5) \\
    $10.0\--25.0$~keV & $0.25\pm0.04$ & $-0.0003\pm0.0005$ & 1.068 (11) & & $0.29\pm0.05$ & $-0.002\pm0.002$ & 0.135 (5)\\
\enddata
\tablecomments{Errors represent 68\% confidence level.}
\end{deluxetable*}

\subsection{Phase-resolved Spectra and Model
Fitting \label{sec:spec_fit}}
In a typical phase-resolved spectral analysis, many compare spectra extracted from limited time intervals around the peak and bottom of the pulsation. For example, \citet{Koliopanos2019}, who analyzed the same data set as the present study, utilized 60\% (30\% each for on and off-pulse spectra) of a rotation cycle and left the rest from the phase-resolved analysis. Although this may clarify the difference between the on-pulse and off-pulse spectra, one can also discard the photon statistics excessively by restricting the time interval, and lead several models to degenerate.

In contrast to the ordinary phase-resolved analysis, our study enables us to add further information to untangle the model degeneracy and utilize the data more efficiently. As the CCPs clarified in the previous section, the entire pulsation cycle of NGC~300 ULX-1 can be categorized into two phase intervals, namely the ``faint phase'' and the ``bright phase''. Since we now visually know from the CCPs that the spectral shape is constant within each phase, we do not have to limit the time interval further to see the change in spectrum. Hence, we hereafter divide the entire observational data into these two phases and extract spectra from each using all of the data therein to study what spectral model composition can explain the spectrum from each epoch with as high photon statistics as possible.

\begin{figure}
    \centering
    \includegraphics[width=\columnwidth]{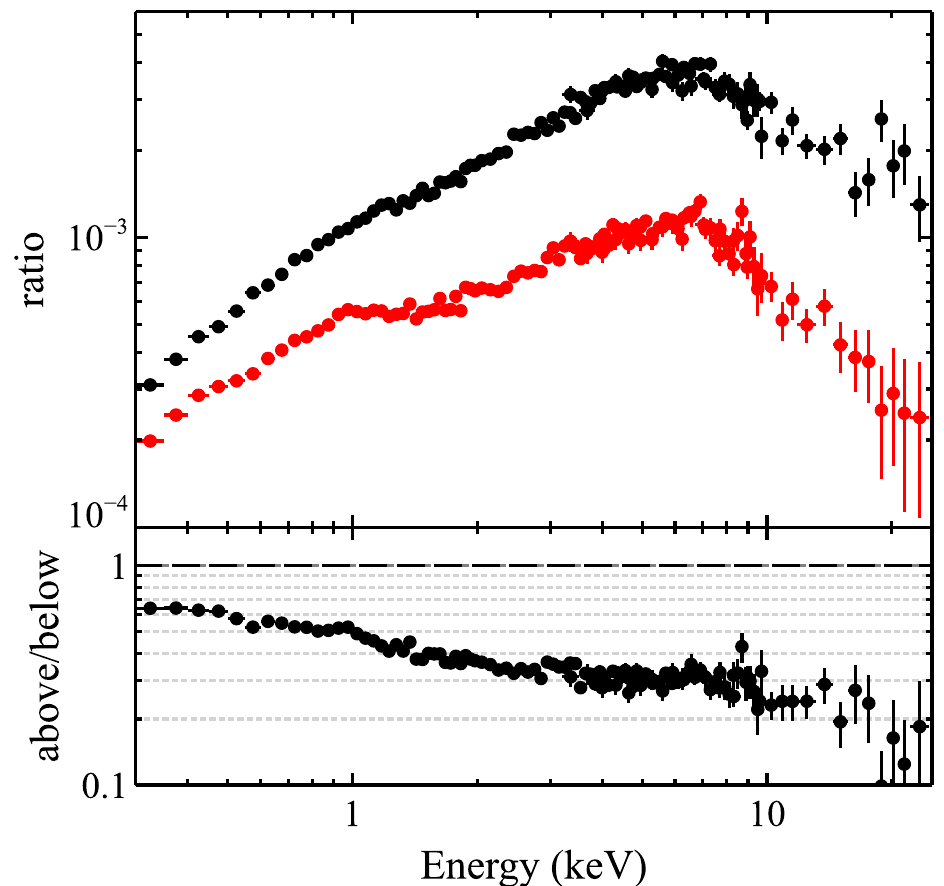}
    \caption{Upper panel; phase-resolved spectra (on phase: black, off phase: red). The instrumental response is tentatively removed by taking ratios over a power-law model with a photon index of 2. Bottom panel; spectral ratio between two phases.}
    \label{fig:plra_below_above}
\end{figure}
We present the bright phase (black) and the faint phase (red) spectra in the top panel of Figure \ref{fig:plra_below_above}. To tentatively unfold the spectra with the instrumental response, they are shown in ratios over a power-law model with a photon index of $\Gamma=2$. Thus, ULX-1 exhibits a hard ($\Gamma\sim 1.4$) and continuum-dominated spectrum throughout the pulsation phase. As the pulsating component decreases its intensity, an additional thermal component peaking at 1~keV becomes more apparent above the continuum. 
While the ratio between the two spectra (the bottom panel of Figure \ref{fig:plra_below_above}) indicates a spectral hardening in the $0.3\--5$~keV band, not as apparent in the $> 5$~keV band. This is consistent with the behavior we saw in the ``breaking'' feature of the CCPs presented in section \ref{sec:c3po}. In the following section, we study the underlying emission compositions of these two spectra along with those newly extracted via the C3PO method.

\subsubsection{Spectral Decomposition With the C3PO Method \label{sec:decomp}}
\begin{figure*}
    \centering
    \includegraphics[width=\textwidth]{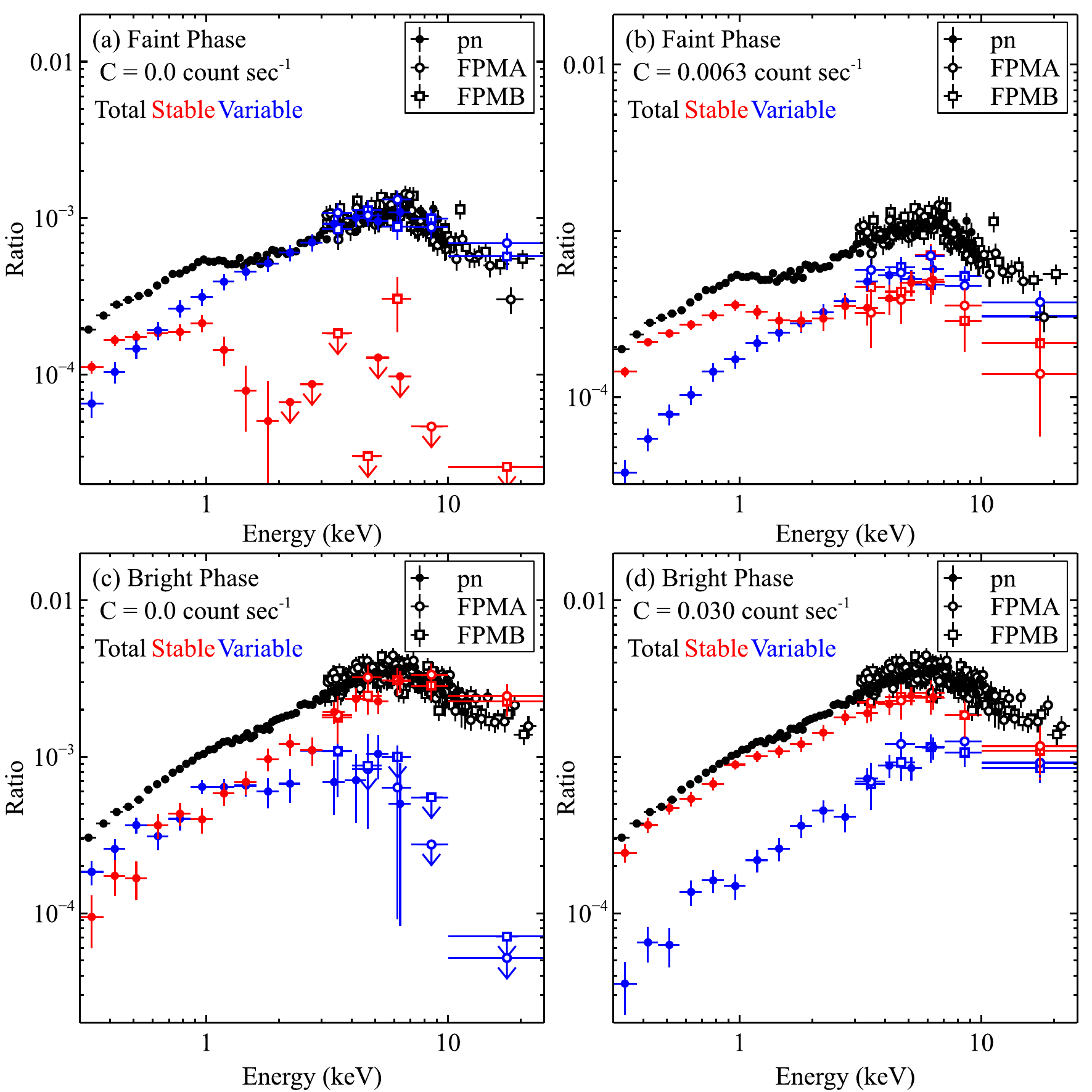}
    \caption{Phase resolved (black) and decomposed stable (red)/variable (blue) component spectra. They are shown in ratios over a power-law model for the same reason as Figure \ref{fig:plra_below_above}. The difference between spectra in left hand and right hand side panels is whether assuming zero (left) or certain count rates (right) as the intensity floor (see text for detail). Filled and open markers represent the data from XMM-Newton and NuSTAR, respectively. Data from NuSTAR FPM-A are shown in circle markers and those from FPM-B are in squares.}
    \label{fig:spec_comp}
\end{figure*}
In Figures \ref{fig:spec_comp} (a) and (c), we present the identical phase-resolved spectra as Figure \ref{fig:plra_below_above} (black) with those newly decomposed via the C3PO method (colored). Thus, we successfully extracted two spectral pairs of the stable and variable components. The former of the faint phase has a convex shape peaking at sub-keV (red in Figure \ref{fig:spec_comp} a), accounting for the noticeable ``soft excess'' mentioned in Figure \ref{fig:plra_below_above}. In addition, it lacks any significant emission at the hard energy band. The FPM-B data showed a finite offset and a different slope from FPM-A by $\sim2\sigma$ at $5.28\--7.00$~keV. This could be interesting if it was real as the energy band includes the peak energy of the Fe K$\alpha$ line. However, XMM-Newton pn and FPM-A, in contrast, gave only upper limits to the stable component at this energy. Considering the five times higher effective area of XMM-Newton and the fact that FPM-A is consistent with XMM-Newton, we conclude that the offset and inconsistent slope in FPM-B is an artifact induced by statistical fluctuation. The spectrum of the variable component (or the pulsating component, in other words) is expanding through a wide energy band of $0.4\--25$~keV with a characteristic rollover at seven $\sim7$~keV (blue in Figure \ref{fig:spec_comp} a). In the bright phase (Figure \ref{fig:spec_comp} c), the stable component (red) exhibits a similar convex spectrum as the faint-phase one except for the extended distribution that drops off sharply at $\sim6$~keV. Whereas, the variable component (blue) yields a similarly hard-bending continuum as that in the faint phase.

The decomposition done above assumes that the count rate of the reference band reaches zero at minimum, which is not necessarily guaranteed. In fact, the pulse fraction of $7.0\--10$~keV is relatively high ($\sim80\%$) but does not reach $100\%$. Since the CCPs in section \ref{sec:c3po} reflects the variability of the pulsating component, we thus may have to treat a particular count rate of $7.0\--10.0$~keV as a net zero point of the analysis to decompose the spectrum into the pulsating accretion flow component from that of non-pulsating one. Therefore, we consider the non-zero case by employing an ``intensity floor'' as \citet{Noda2014} did.

To account for the intensity floor, we shift equation \ref{eq:linear} toward the $+x$ direction by the floor count rate of $c$. Hence equation \ref{eq:linear} can be rewritten as
\begin{equation}
    y=a(x-c)+b',
    \label{eq:shift_linear}
\end{equation}
where
\begin{equation}
    b' = b+ac.
    \label{eq:shift_offset}
\end{equation}
Here, we employed the floor count rate equivalent to the minimum count rate in each phase, namely $c=0.0063$~count~sec$^{-1}$ for the faint phase and $c=0.030$~count~sec$^{-1}$ for the bright phase (e.g., see \ref{fig:ccplot}). The revised spectra of both phases are shown in Figure \ref{fig:spec_comp} (b) and (d). As can be followed from equation \ref{eq:shift_linear} and \ref{eq:shift_offset}, the revision changes the shape and normalization of the stable-component spectrum. As for variable one, however, it only scales the normalization by a factor of $1-c/x_{0}$ and does not affect its spectral shape.

The revised stable-component spectrum is generally a summation of that in the $c=0.0$ case and a fraction of the variable component. In fact, the spectral shapes of the revised two components in the faint phase are nearly identical at $>2$~keV (Figure \ref{fig:spec_comp} b). We presume that the pulsating component is still present in the spectrum at the pulse minimum, and it is accounting for the deficient $20\%$ of the count rate for the pulse fraction to reach $100\%$. Hence, to separately discuss the pulsating-component spectrum and that of the non-pulsating one, we hereafter assume the $c=0.0$~count~sec$^{-1}$ case in the analysis of faint phase spectra.

Generally, the floor intensity in the C3PO method is the lower limit down to where we can safely extrapolate the variability. Although we assumed $c=0.0$~count~sec$^{-1}$ in the faint phase, this may not be appropriate in the other from this perspective because the spectral mode shifts rather continuously from one to another, which is apparent in Figure \ref{fig:brek_vs_power}, and employing $c=0.0$~count~sec$^{-1}$ in the bright phase will drop this information. As a matter of fact, the spectral shape of the stable component in Figure \ref{fig:spec_comp} (c) is inconsistent with that of the faint phase. This is due to assuming that the variable component can decrease its intensity to zero, and it is not the case in this particular phase. Since the bright phase mode contains a solid floor intensity, namely the break count rate, we thus employ the $c=0.030$ count~sec$^{-1}$ case in the following analysis.

\subsubsection{Model Fitting of the Faint-phase Spectra \label{sec:fit_faint}}
\begin{figure}
    \centering
    \includegraphics[width=\columnwidth]{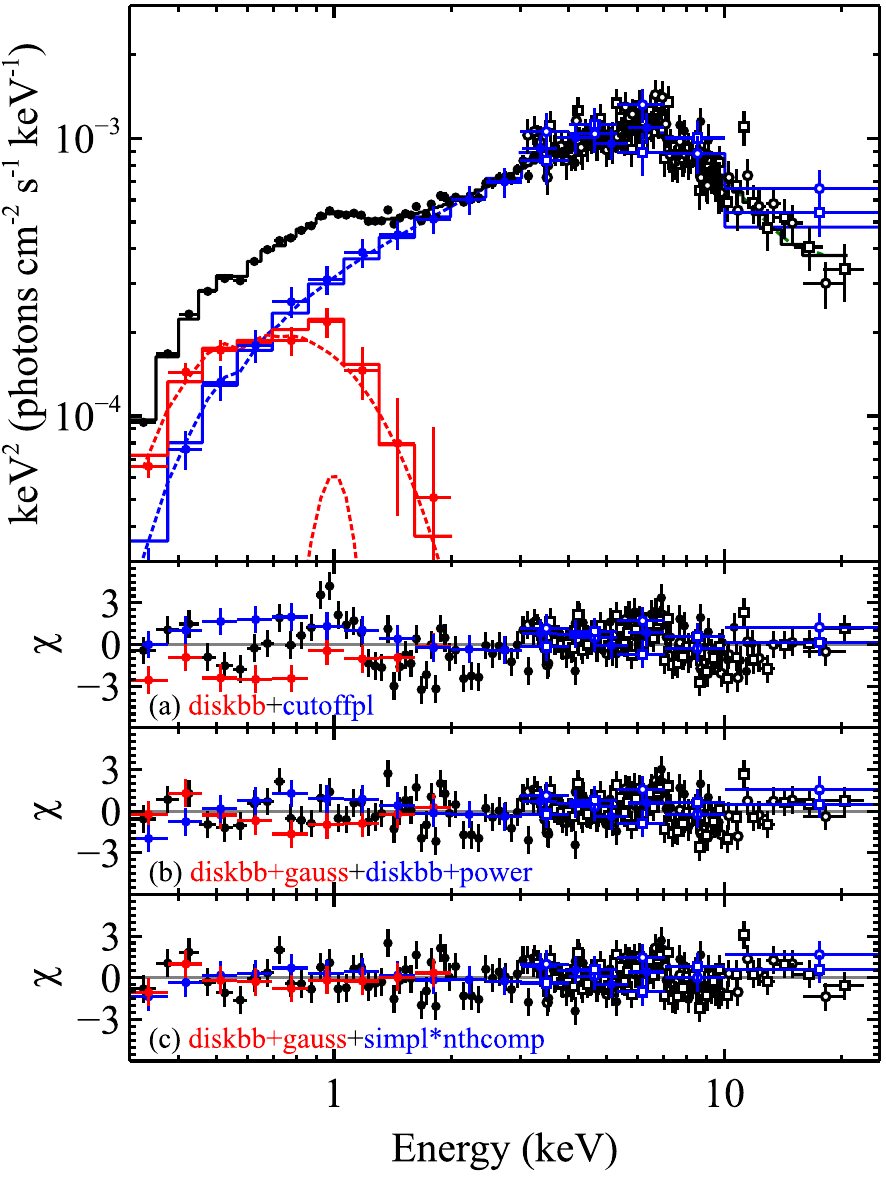}
    \caption{Top panel; spectra and the best-fit model unfolded with the instrumental response. The faint phase, stable component, and variable component spectra are shown in black, red, and blue, respectively. Panels (a), (b), and (c); residuals from \texttt{diskbb$_{\tt st}+$cutoffpl$_{\tt v}$}, \texttt{diskbb$_{\tt st}+$gauss$_{\tt st}+$diskbb$_{\tt v}+$power$_{\tt v}$}, and \texttt{diskbb$_{\tt st}+$gauss$_{\tt st}+$simpl$_{\tt v}*$nthcomp$_{\tt v}$}, respectively}.
    \label{fig:below_model_fit}
\end{figure}

We begin by fitting the spectrum extracted from events in the faint phase. The original (black), stable component (red), and variable component spectra (blue) are presented in the uppermost panel of Figure \ref{fig:below_model_fit}. As we thus have successfully decomposed the spectrum of NGC~300~ULX-1 into the pulsating and stable components, we fit these three spectra simultaneously with a pair of emission models that each represents the individual components.

Since the emission mechanism of super-critical accretion flow is still poorly understood, physically reasonable spectral modeling is yet to be established for the pulsating emission component of ULXPs. Hence, studies on the component usually rely on models relatively empirical. Reflecting its hard and extending shape continuum with a characteristic rollover at $\sim 7$~keV, the most frequently seen example of the modeling is one utilizing a power law with an exponential cutoff (\texttt{cutoffpl} in XSPEC expression). In fact, \citet{Brightman2016a} and \citet{Walton2018a} successfully reproduced the hard continuum of M82 X-2 and NGC~7793 P-13 with this model, respectively. On the other hand, the parameters in the \texttt{cutoffpl} model have no physical meanings. This has motivated some authors (e.g., \citealt{Koliopanos2019}) to alternatively use a multi-color disk blackbody model (\texttt{diskbb} in XSPEC expression; \citealt{Mitsuda1984}) or Comptonization model (\texttt{nthcomp} in XSPEC expression; \citealt{Zdziarski1996}) to approximate the emission physics expected from several theoretical studies of super-critical accretion flows (e.g., \citealt{Mushtukov2017}). In the present paper, we test these three patterns of modeling to explain the pulsating component that dominates the hard energy band.

 As for the stable component, we employ either a single-temperature blackbody or a multi-color disk blackbody model. This is because the emission has a characteristic convex shape that is likely to be an optically thick thermal emission. In addition, it clearly originates from a region formed somewhere around the neutron star. Even if the accretion disk is the origin of the stable component, it is still unclear whether the disk is in the standard accretion state derived by \citet{Shakura1973}. Therefore, we test two patterns of disk blackbody models, one assuming the disk to be in the standard accretion regime (\texttt{diskbb} in XSPEC; \citealt{Mitsuda1984}) and the other having a radial temperature dependence as an additional free parameter (\texttt{diskpbb} in XSPEC).

Finally, we commonly multiply a photoelectric absorption model, \texttt{tbabs}, and a constant factor to both components. The former is to take the absorption below 1~keV into account and has a column density as a free parameter. It represents the hydrogen-equivalent number density of matter within the line of sight, which includes absorption by the Galactic interstellar medium and intrinsic within the NGC 300 galaxy. We assumed the solar abundance and used a material table by \citet{Wilms2001}. The latter is to account for the systematic errors in absolute count rate due to using spectra from different instruments. Since models with the same names are used to reproduce distinct components, we hereafter clarify to which the model belongs by denoting the one explaining the stable one with an \texttt{st} subscript and the other with a \texttt{v} subscript.

\begin{deluxetable*}{lccccccc}
\tablenum{2}
\tablecaption{The best-fit parameters obtained from the faint phase spectra \label{tab:fit_result}}
\tablewidth{0pt}
\tablehead{
\colhead{model$_{\rm stable}$} & \multicolumn{3}{c}{\texttt{gauss+bb}} & & \multicolumn{3}{c}{\texttt{gauss+diskbb}}\\
\cline{2-4} \cline{6-8}
\colhead{model$_{\rm variable}$} & \colhead{\texttt{diskbb+power}} & \colhead{\texttt{simpl*cutoffpl}} & \colhead{\texttt{simpl*nthcomp}} & & \colhead{\texttt{diskbb+power}} & \colhead{\texttt{simpl*cutoffpl}} & \colhead{\texttt{simpl*nthcomp}}
}
\startdata
    & \multicolumn{7}{c}{stable component} \\
    $T_{\rm in/bb}^{a}$ (keV) & $0.19\pm0.01$ & $0.190\pm0.003$ & $0.18\pm0.01$ & & $0.25\pm0.02$ & $0.24^{+0.04}_{-0.02}$ & $0.25^{+0.02}_{-0.03}$\\
    norm$_{\rm disk}^{b}$ & - & - & - & & $12^{+6}_{-4}$ & $12.8\pm0.4$ & $10^{+7}_{-3}$\\
    norm$_{\rm bb}^{c}$ ($\times10^{-6}$)& $6.9\pm0.6$ & $6.40^{+0.1}_{-0.5}$ & $5.7^{+0.5}_{-0.4}$ & & - & - & - \\
    $E_{\rm line}^{d}$ (keV) & $1.00\pm0.04$ & $1.00\pm0.03$ & $1.00\pm0.03$ & & $0.94^{+0.05}_{-0.07}$ & $0.94\pm0.02$ & $0.97^{+0.04}_{-0.05}$ \\
    $\sigma^{e}$ (keV) & $<0.1$ & $<0.1$ & $<0.1$ & & $0.15^{+0.08}_{-0.06}$ & $0.29\pm0.02$ & $0.12^{+0.07}_{-0.05}$\\
    \hline
     & \multicolumn{7}{c}{variable component} \\
    $\Gamma_{\rm simpl}^{f}$ & - & $1.00^{+0.17}_{-0.02}$ & $1.0^{+0.5}_{-0.2}$ & & - & $1.03^{+0.25}_{-0.02}$ & $1.2^{+0.4}_{-0.2}$ \\
    $F^{g}$ & - & $0.40^{+0.4}_{-0.02}$ & $0.20^{+0.18}_{-0.06}$ & & - & $0.40\pm0.02$ & $0.22^{+0.19}_{-0.06}$\\
    $T_{\rm in/bb}^{a}$ (keV) & $2.34\pm0.08$ & - & $0.21\pm0.02$ & & $2.42\pm0.09$ & - & $0.14\pm0.04$\\
    norm$_{\rm disk}^{b}$ ($\times10^{-3}$) & $4.2\pm0.5$ & - & - & & $3.6\pm0.5$ & - & -\\
    $T_{\rm cut/e}^{h}$ (keV) & - & $4.44^{+0.03}_{-0.05}$ & $1.77\pm0.09$ & & - & $4.32^{+0.2}_{-0.05}$ & $1.73\pm0.09$\\
    $\Gamma_{\rm pl/nthcomp}^{i}$ & $1.7\pm0.1$ & $0.70\pm0.05$ & $1.57\pm0.03$ & & $1.9\pm0.1$ & $0.81\pm0.05$ & $1.56\pm0.04$\\
    norm$^{j}_{\rm pl/nthcomp}$ ($\times10^{-4}$) & $1.1\pm0.3$ & $6.24^{+0.31}_{-0.06}$ & $3.9^{+1.2}_{-0.4}$ & & $1.6\pm0.3$ & $6.46\pm0.06$ & $4.1^{+1.0}_{-0.5}$\\
    \hline
     & \multicolumn{7}{c}{common component} \\
    $N_{\rm H}^{k}$ ($\times10^{20}$~cm$^{-2}$) & $3\pm1$ & $1.7\pm0.8$ & $1\pm1$ & & $6\pm1$ & $4.6\pm0.2$ & $5\pm1$\\
    $\chi^2/\nu^{l}$ & 265.38/205 & 287.24/206 & 247.60/203 & & 248.80/205 & 287.7/206 & 219.87/203\\
\enddata
\tablecomments{a: Temperature of the inner-disk radius or blackbody surface. b: Normalization parameter of the \texttt{diskbb} model. c: Normalization parameter of the \texttt{bb} model. d: Center energy of the Gaussian line. e: Standard deviation of the Gaussian line. f: Photon index of the power-law continuum that \texttt{simpl} creates. g: Fraction of the photons that is devoted to create the power-law continuum by \texttt{simpl}. h: Cutoff energy of \texttt{cutoffpl} or temperature of the Comptonizing electron cloud. i: Photon index of the \texttt{cutoffpl} model or that of the \texttt{nthcompl}. j: Normalization parameter of \texttt{cutoffpl} or \texttt{nthcomp}. Represents the photon flux at 1~keV. k: Hydrogen-equivalent column density. l: Chi-squared statistics and degree of freedom.}
\end{deluxetable*}
Despite utilizing models that were successful in several previous ULX studies, none of the patterns gave acceptable fits to the spectra. As a representative, we present residuals between the \texttt{diskbb$_{\tt st}$+cutoffpl$_{\tt v}$} model and data in panel (a) of Figure \ref{fig:below_model_fit}. We can confirm a significant enhancement at $\sim 1$~keV and a slight wiggling feature that goes downwards in $7\--10$~keV and then upwards in $>10$~keV. Furthermore, the entire stable component model overestimates the data, which we discuss later. 

The residual in $\sim 1$~keV is likely to be an emission-line-like feature occasionally reported from several ULXs including ULXPs (for non-pulsating ULXs e.g., NGC~5408 X-1, NGC~6946 X-1 \citealt{Middleton2014}, NGC~1313 X-1 \citealt{Pinto2016}, and NGC~247 ULX-1 \citealt{Pinto2021}; for ULXPs e.g., SMC X-3 \citealt{Koliopanos2019}). The feature is also reported in a different observation of ULX-1 \citep{Ng2022}, and considered to be Fe L, Ne X, or Fe XVIII emission lines from the surrounding gas. Since the enhancement is also present in the stable component spectrum, we hence modify the model for the component by adding a Gaussian line at $\sim 1.0$~keV.

The wiggling residual is also widely seen in the hard X-ray band of ULXs (e.g., \citealt{Bachetti2013}, \citealt{Walton2018a}), indicating that spectra require an extra component that extends up to $\sim 25$~keV. A simple solution for this is to introduce another power law (\texttt{power} in XSPEC) to the variable component model. In fact, \citet{Koliopanos2019} have successfully reproduced the $> 2$~keV spectrum of this ULXP, ULX-1, with a combination of multi-color disk and a power-law model. As another option, it is also effective to multiply a model that modifies a part of the current model shape to an extending power law. For example, \citet{Walton2018b} and \citet{Walton2014} multiplied the \texttt{simpl} model to a cutoff power law and a Comptonization model, respectively. The model redistributes a part of the photons from the multiplied model to form an additional power-law emission in higher energy, which technically gives similar effects as adding a power-law model. The difference from adding a power law is that the \texttt{simpl} model extends only toward higher energy, whereas the power law does to infinity in both energy directions. Therefore, \texttt{simpl} avoids photon numbers diverting in the lower energy band. In the following analysis, we refit the variable component with these modified models, namely \texttt{power$_{\tt v}$}$+$\texttt{diskbb$_{\tt v}$}, \texttt{simpl$_{\tt v}$}$*$\texttt{cutoffpl$_{\tt v}$}, and \texttt{simpl$_{\tt v}$}$*$\texttt{nthcomp$_{\tt v}$}.

The parameters and the goodness of fit are summarized in Table \ref{tab:fit_result}, and the best-fit model and examples of the residuals are shown in Figures \ref{fig:below_model_fit} (b), and (c), respectively. Let us begin by comparing the results in terms of the difference in variable component models. Although the discrepancy in $\sim1$~keV and $>10$~keV are significantly improved, the models including \texttt{cutoffpl$_{\tt v}$} for the variable component failed to simultaneously reproduce the stable component just like the one previously shown in Figure \ref{fig:below_model_fit} (a). This is mainly because the \texttt{cutoffpl$_{\tt v}$} model is causing a clash with the stable component model due to its extending characteristic mentioned in the previous paragraph. It is also the same in the model that includes \texttt{diskbb$_{\tt v}$}$+$\texttt{power$_{\tt v}$} in it. Since the photons of the \texttt{diskbb$_{\tt v}$} model quickly drop off in lower energy due to the Rayleigh-Jeans region, the model gave a better fit than those using \texttt{cutoffpl$_{\tt v}$} (Table \ref{tab:fit_result}). Still, the \texttt{power$_{\tt v}$} model introduced to account for the $>10$~keV causes a conflict with the stable component at $0.5\--1.3$~keV shown in Figure \ref{fig:below_model_fit} (b). 

In contrast to the present work, the \texttt{diskbb$_{\tt v}$}$+$\texttt{power$_{\tt v}$} model was successful to explain the hard X-ray spectrum in the previous work using the same data set \citep{Koliopanos2019}. In the analysis, authors relied only on the phase-resolved spectra like the ordinary phase-resolving spectral analysis, and no other data were available to restrict the shape of the models that form the lower energy continuum in particular. Such lack of restriction can allow \texttt{tbabs} and the cooler \texttt{diskbb$_{\tt st}$ (or \texttt{bb$_{\tt st}$})} model to absorb the overshooting power-law component by increasing the column density and/or decreasing its normalization. \citet{Koliopanos2019}, in fact, pointed out that a strong correlation was present between the column density and the power-law photon index. Hence, the models have degenerated in the \citet{Koliopanos2019} case, and thanks to the C3PO method, we have successfully resolved this by providing the models additional spectra to reproduce simultaneously with the phase-resolved one.

The radial surface temperature of the accretion disk model $T(r)$ is proportional to a negative power of the disk radius $r$, namely $T(r)\propto r^{-p}$, where $p$ is the power index. In the \texttt{diskbb} case, $p$ is fixed to 0.75 with which \citep{Koliopanos2019} approximated the temperature gradient in the pulsating accretion flow. To test whether the fit improves by letting this $p$ vary, we replaced the \texttt{diskbb$_{\tt v}$} model in the variable component with the \texttt{diskpbb$_{\tt v}$} model that has the power index $p$ as an additional free parameter. However, letting $p$ vary did not improve the fit; it gave only a 3.5 smaller chi-squared value for one degree of freedom less than the \texttt{diskbb$_{\tt v}$} case with $p=0.95$. As a result, the models utilizing \texttt{nthcomp$_{\tt v}$} and \texttt{simpl$_{\tt v}$}, which yields fewer photons in lower energy due to the Rayleigh-Jeans break of \texttt{nthcomp$_{\tt v}$} and the non-diverging characteristic of \texttt{simpl$_{\tt v}$}, gave the best fit among the variable component model patterns as shown in Figure \ref{fig:below_model_fit} (c).

Instead of creating an extending power-law component, \citet{Walton2018a} argued that they successfully reproduced the residuals in $> 10$~keV with a wide absorption line possibly originating from a cyclotron resonance scattering feature. We also tested this alternative solution by multiplying an absorption line model (\texttt{gabs} in XSPEC expression) to the continuum models. Since the photon statistics at $>10$~keV are too poor to determine the line central energy, we fixed the value to that of \citet{Walton2018a}; 12.8~keV. The model gave the same tendency as those shown above. Only the model utilizing \texttt{nthcomp$_{\tt v}$} gave an acceptable fit, which is rather natural because \texttt{gabs} affects only the spectral shape above 10~keV. The best fit was slightly better ($\chi^2/\nu=247.26/217$) but insufficient to rule out the model using \texttt{simpl$_{\tt v}$}. The obtained line width of $4.2^{+1.1}_{-0.6}$~keV is consistent with that in \citet{Walton2018a}; $3.1^{+0.8}_{-0.7}$~keV. The model exhibited the same parameter values as those in the model using \texttt{simpl$_{\tt v}$} within the errors, except for a higher electron temperature ($2.7\pm0.1$~keV) and a slightly harder photon index ($1.51^{+0.05}_{-0.07}$). Hence, we conclude at least the variable component requires a model that breaks sharply in the lower energy end and cannot distinguish whether an extending power law or an absorption line is the best model to account for the spectral feature above 10~keV.

We next compare the models used to explain the stable component. All the models utilizing \texttt{diskbb$_{\tt st}$} gave a slightly better fit than those including \texttt{bb$_{\tt st}$}. This is because \texttt{diskbb$_{\tt st}$} yields a softer spectrum than \texttt{bb$_{\tt st}$} due to the contribution of cooler blackbody emission from the outer disk region. The residual indicates that the stable component spectrum is too wide to be explained with the later model. Furthermore, its hard spectral nature forces the \texttt{bb$_{\tt st}$} model to compensate for its lack of photon in the lower energy band by making the absorption column density smaller ($N_{\rm H}< 3.6\times10^{20}$~cm$^{-2}$) than those of \texttt{diskbb$_{\tt st}$} (Table \ref{tab:fit_result}). In particular, that of the best-fit pattern being comparable to or even smaller than the Galactic value ($\sim 2\times 10^{20}$~cm$^{-2}$; \citealt{Dickey1990}), and we consider this unreasonable. Gathering these results together, we conclude that the data favor the \texttt{diskbb$_{\tt st}$} model for the stable component spectrum.

To test whether the multi-color disk blackbody emission differs from the standard disk, we also allowed the radial disk temperature dependency to vary by replacing \texttt{diskbb$_{\tt st}$} with \texttt{diskpbb$_{\tt st}$} as we did in the variable component modeling. While the temperature profile is $p=0.75$ in the standard accretion disk \citep{Shakura1973}, it is expected to be flatter if the disk deviates from the standard regime as the accretion rate increases. In a near-Eddington accretion rate, the disk is expected to reach a state called a slim disk, in which the disk has radial temperature with $p=0.5$ \citep{Watarai2000, Watarai2001}. The best fit for $p$ is $0.63$, which is in the middle between the slim disk and the standard disk. However, allowing $p$ to vary did not improve the fit, giving only $\Delta\chi^2=0.5$ with -1 degree of freedom from the model assuming the standard disk ($p=0.75$). Fixing p to 0.5, namely assuming the slim disk, still gave nearly identical goodness of fit ($\Delta\chi^2=1.0$) with a higher column density of $\sim6\times10^{20}$~cm$^{-2}$ than those in the standard disk case. This is because the slim disk has a softer spectrum than the standard disk due to its flatter temperature gradient, forcing the absorption model to give higher column density. Hence, we cannot statistically distinguish whether the stable component favors the slim disk state or the standard accretion regime.

Although we thus have successfully reproduced the entire continuum component of ULX-1, a narrow positive excess is still present at $\sim 6.7$~keV, which is consistent with the energy of the He-like Fe K$\alpha$ line. In fact, adding a Gaussian line improved the fit by $\Delta\chi^2=20.6$ with a decrease of 3 degrees of freedom. Since the upper limits on the count rate of the stable component are well below the expected strength of the line at this energy, we here assumed that the possible emission line feature belongs to the variable one. The best-fit center energy and width of the Gaussian line were $6.7^{+0.1}_{-0.4}$~keV and $<0.6$~keV, respectively.

To evaluate the statistical significance of this line-like feature, we generated 5000 simulated spectra based on the previous best-fit model that only consists of a continuum around that energy and tested how further the fit improves by adding a Gaussian line at $6.7$~keV on each spectrum. The simulated spectra are created with HEASOFT \texttt{fakeit} command, which generates response-folded spectra of a given spectral model with expected Poisson noise. The exposure of each spectrum is set to be equivalent to the actual observation. To prevent the fit to diverge, we limited the central energy and the width of the line to vary within $6.0\--7.0$~keV and $< 1.0$~keV, respectively.

\begin{figure}
    \centering
    \includegraphics[width=\columnwidth]{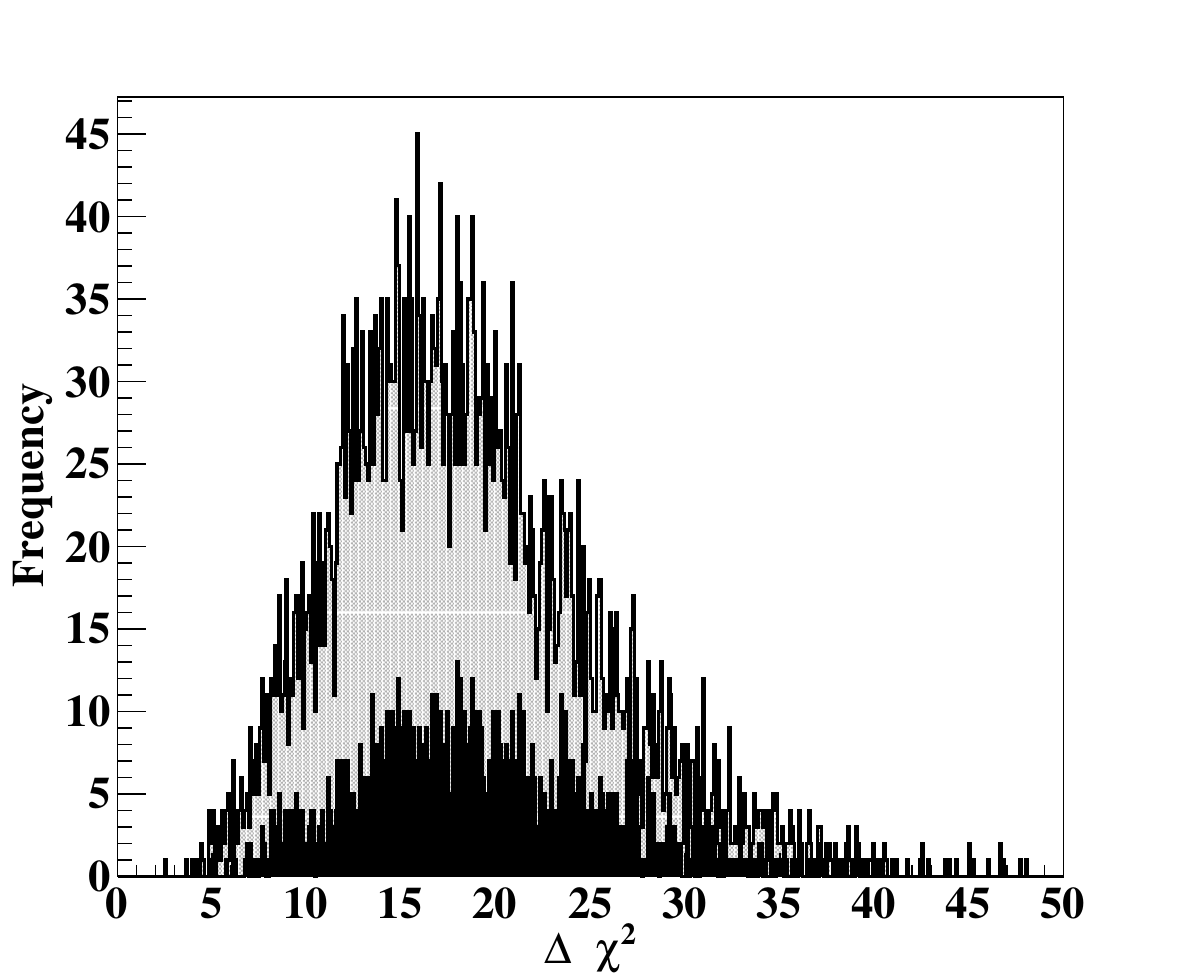}
    \caption{The $\Delta\chi^2$ distribution obtained from adding a Gaussian line to the 5000 simulated spectral fits. The grey-hatched histogram represents the distribution of the total simulations, and the filled-black one is for those gave values consistent with the observational results within the statistical errors.}
    \label{fig:delchi_dist}
\end{figure}

Figure \ref{fig:delchi_dist} presents the distributions of simulated spectral fit improvements in terms of a difference in $\chi^2$ statistics ($\Delta\chi^2$) between the two spectral model fittings. One is our best-fit model (\texttt{diskbb$_{\tt st}$+gauss$_{\tt st}$+simpl$_{\tt v}$*nthcomp$_{\tt v}$}), and the other is that with an additional Gaussian line at $\sim6.7$~keV. The simulated spectra that gave better improvement than the observational result ($\Delta\chi^2>20.6$) are $\sim40\%$ of the total simulation (grey histogram). Furthermore, if we limit to those exhibited consistent parameters as the observation within the $90\%$ significance error (black histogram), the number decreases to $9.6\%$. Therefore, we conclude that the statistical significance level of the possible line feature is $\sim90\%$, which is rather promising but insignificant to claim as a detection.

Considering that an Fe K$\alpha$ line is present in a spectrum of ULXP SMC~X-3 \citet{Koliopanos2018}, it is natural to expect the same for NGC~300~ULX-1, and the possible feature we observed above is a good candidate. In addition to insufficient photon statistics in the total spectrum, the present detectors do not have enough effective area to divide those obtained from the C3PO method into finer bins, which are making hard to resolve the feature in the variable component spectrum. To strengthen the statistical significance and confirm the origin of this feature, we strongly advise observing this source with a long exposure or observatories having a higher effective area, such as NICER, and conducting the same analysis on that data set.

Finally, to test how the assumption made in section \ref{sec:decomp} can affect our result, we performed model fittings using the spectral set assuming a non-zero floor intensity of $C=0.0063$~count~sec$^{-1}$ (the spectra in Figure \ref{fig:spec_comp} b). As described in section \ref{sec:decomp}, the floor intensity does not change the shape of the variable component spectrum. In addition, the stable component spectrum becomes a summation of the $C=0.0$ case and some fractions of the variable component. Accordingly, we used the same models as those we have employed so far, except for the variable component model being added to the stable component one. The results are relatively the same as those we have confirmed in our previous analysis. The best-fit model is \texttt{diskbb$_{\tt st}$+gauss$_{\tt st}$+simpl$_{\tt v}$*nthcomp$_{\tt v}$}, and the obtained parameters are consistent with those in Table \ref{tab:fit_result} within the errors. Models using \texttt{cutoffpl$_{\tt v}$} or \texttt{power$_{\tt v}$} caused conflicts with the stable component, and those using \texttt{bb$_{\tt st}$} to explain the soft excess exhibited worse fit or column density smaller than the Galactic value. Hence we conclude that the assumption made on the floor intensity does not affect our results.

\subsubsection{Model Fitting of the Bright Phase Spectra}
\begin{figure}
    \centering
    \includegraphics[width=\columnwidth]{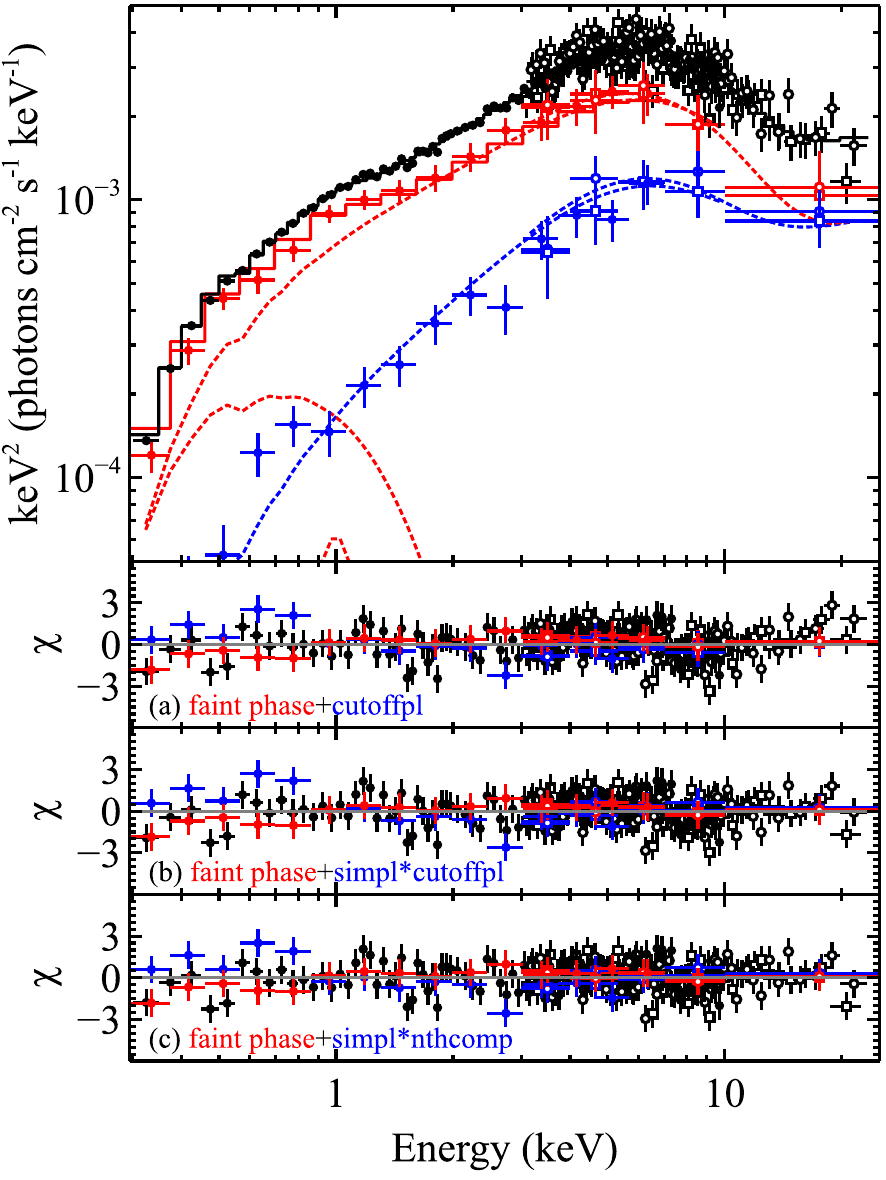}
    \caption{Top panel; spectra and the best-fit model unfolded with the instrumental response. The bright phase, stable component, and variable component spectra are shown in black, red, and blue , respectively. Panels (a), (b), and (c); residuals from the best-fit faint phase model$+$, \texttt{cutoffpl$_{\tt v}$}, \texttt{simpl$_{\tt v}*$cutoffpl$_{\tt v}$}, and \texttt{simpl$_{\tt v}*$nthcomp$_{\tt v}$}, respectively.}
    \label{fig:above_model_fit}
\end{figure}

The spectra of the bright phase, stable component, and variable component are shown in the top panel of Figure \ref{fig:above_model_fit} in black, red, and blue, respectively.
Although the errors are relatively large due to the limited data points in the CCPs (Figure \ref{fig:ccplot}), the stable component exhibits the characteristic ``two-humped'' spectrum similar to that we saw in the faint phase. As for the variable component spectrum, it shows a slightly harder continuum than the stable component peaking at $\sim7$~keV. 

Here, we take the results from the faint phase into account. Since the CCPs in section \ref{sec:c3po} were continuous at the breaking feature, the spectral shift from the faint phase to the bright phase should also be smoothly connected. Therefore, we assume the model gave the best fit in the faint phase spectra (\texttt{diskbb$_{\tt st}$+gauss$_{\tt st}$+simpl$_{\tt v}$*nthcomp$_{\tt v}$}) for the stable component of this phase. Since the shape of the variable component spectrum does not change within the faint phase, we fixed all the parameters to the best-fit values shown in Table \ref{tab:fit_result} except for the normalization of \texttt{simpl$_{\tt v}$*nthcomp$_{\tt v}$}. 

\begin{deluxetable*}{lccccc}
\tablenum{3}
\tablecaption{The best-fit parameters obtained from the spectra in the bright phase. \label{tab:fit_result_above}}
\tablewidth{0pt}
\tablehead{
\colhead{variable component model} & \colhead{\texttt{diskbb}} & \colhead{\texttt{cutoffpl}} & \colhead{\texttt{diskpbb+power}} & \colhead{\texttt{simpl*cutoffpl}} & \colhead{\texttt{simpl*nthcomp}}
}
\startdata
    norm$_{\rm nthcomp}$ ($\times10^{-4}$) & $8.5\pm0.2$ & $8.7\pm0.2$ & $8.6\pm0.2$ & $8.5\pm0.2$ & $8.5\pm0.1$\\
    $\Gamma_{\rm simpl}$ & - & - & - &  $1.2^{+0.9}_{-0.2}$ & $1.7^{+0.3}_{-0.2}$\\
    $F$ & - & - & - & $0.4\pm0.2$ & $0.38\pm0.04$\\
    $T_{\rm in/bb}$ (keV) & $3.2\pm0.1$ & - & $2.4\pm0.3$ & - & $0.22\pm0.02$\\
    $p^{a}$ & - & - & $> 0.8$ & - & - \\
    $T_{\rm cut/e}$ (keV) & - & $4.8^{+0.5}_{-0.4}$ & - & $3.3^{+0.4}_{-0.5}$ & $1.5\pm0.2$\\
    $\Gamma_{\rm pl/nthcomp}$ & - & $0.40\pm0.08$ & $0.9^{+0.3}_{-1.6}$ & $0.2\pm0.1$ & $1.36\pm0.01$\\
    norm$_{\rm pl/nthcomp}$ ($\times10^{-4}$) & - & $2.1\pm0.2$ & $0.2^{+0.5}_{-0.2}$ & $3.3^{+1}_{-0.7}$ & $2.3\pm0.1$\\
    norm$_{\rm disk}$ ($\times10^{-4}$) & $16\pm3$ & - & $60^{+50}_{-20}$ & - & - \\
    $\chi^2/\nu$ & $316.11/292$ & $299.21/291$ & $280.70/289$ & $280.88/289$ & $269.23/288$\\
\enddata
\tablecomments{a: The power index of the radial temperature profile of the disk. The rest are the same as those in Table \ref{tab:fit_result}}
\end{deluxetable*}
The variable component has a continuum extending in a power-law manner with a cutoff around $\sim7$~keV as the faint phase one. Hence, we tested the same models as those we used to explain the variable component in section \ref{sec:fit_faint}. The model fitting results are presented in Table \ref{tab:fit_result_above}, and examples of the residual are in Figures \ref{fig:above_model_fit} (a), (b) and (c). Since the variable component gently bends at $\sim7$~keV, neither \texttt{diskbb$_{\tt v}$} nor \texttt{cutoffpl$_{\tt v}$} reproduced the spectrum in $>10$~keV as shown in Figure \ref{fig:above_model_fit} (a). Especially \texttt{diskbb$_{\tt v}$}, of which spectrum sharply drops off with Wien's law at higher energy, gave the worst fit among the models. To reproduce the gradual bent above 10 keV, we again tested the same three modified model combinations for the variable component as section \ref{sec:fit_faint}; \texttt{diskbb$_{\tt v}$+power$_{\tt v}$}, \texttt{simpl$_{\tt v}$*cutoffpl$_{\tt v}$}, and \texttt{simpl$_{\tt v}$*nthcomp$_{\tt v}$}.

Despite adding an extra component at higher energy, \texttt{diskbb$_{\tt v}$+power$_{\tt v}$} gave a similar goodness of fit, $\chi^2/\nu=300.50/290$, as \texttt{cutoffpl$_{\tt v}$} ($\chi^2/\nu=299.21/291$). This is due to the steep $1\--7$~keV continuum, which is too hard to reproduce with the temperature gradient of \texttt{diskbb$_{\tt v}$}. Hence, we again let the gradient, namely the spectral hardness, vary by replacing \texttt{diskbb$_{\tt v}$} with \texttt{diskpbb$_{\tt v}$}. The fit significantly improved as $\chi^2/\nu=280.70/289$ by steepening the temperature gradient $p$ (see Table \ref{tab:fit_result_above}). The fit similarly improved for the rest of the patterns. Especially, The model using \texttt{nthcomp$_{\tt v}$} gave the best fit among them for the same reason as that we mentioned in the faint phase. The dropping-off characteristic of \texttt{nthcomp$_{\tt v}$} is avoiding conflict with the stable component spectrum.

Although we cannot statistically rule out either \texttt{diskpbb$_{\tt v}$+power$_{\tt v}$} or \texttt{simpl$_{\tt v}$*cutoffpl$_{\tt v}$} for the variable component modeling, we hereafter adopt \texttt{simpl$_{\tt v}$*nthcomp$_{\tt v}$} to compare the parameters with those in the faint phase.
\begin{figure}
    \centering
    \includegraphics[width=\columnwidth]{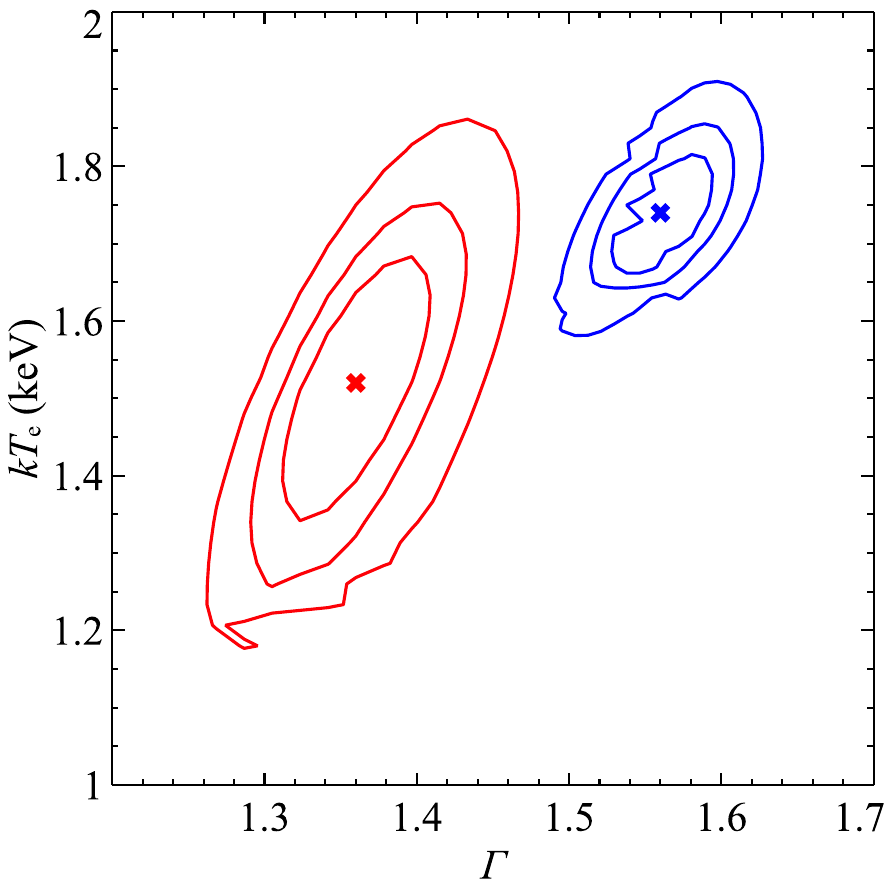}
    \caption{Significance contours of the electron temperature vs photon index of \texttt{nthcomp$_{\tt v}$}. The confidence levels from the faint phase are shown in blue, while that from the bright phase are in red. Inner, middle, and outer solid lines represent 68\%, 90\%, 99.7\% confidence level, respectively. The crosses indicate the best-fit values.}
    \label{fig:contour}
\end{figure}
 It is noticeable that the best fit values between both phases are equivalent within the errors (see Table \ref{tab:fit_result} and Table \ref{tab:fit_result_above}), except for the photon index of \texttt{nthcomp$_{\tt v}$}. The model suggests that the continuum of the variable component is slightly hard in the bright phase. The difference is rather marginal, and we must note that the Comptonization model often exhibits a strong correlation between its electron temperature and photon index. Hence, the errors can be underestimated to some extent. 
 
 To check whether the difference in the photon index is statistically significant, we created a confidence contour of these two (Figure \ref{fig:contour}). The confidence contours do not overlap in the photon index direction with more than a 99.7\% confidence level, from which we can confirm that the continuum of the variable component in this phase is actually harder than the other. The result is consistent with the hint of hardening that was present in Figure \ref{fig:plra_below_above}. On the other hand, the previous study by \citet{Koliopanos2019} did not find such a hardening. We consider the reason is the considerably narrow phase intervals they applied to extract on-pulse and off-pulse spectra. The spectral statistics in the previous study were compromised to clarify the difference between the two and were insufficient to spot such a marginal hardening. Thus, the C3PO method enabled us to maximize the usable spectral statistics and find new features that were not apparent before the present study.

\section{Discussion}
The effective area of XMM-Newton and the high energy of capability of NuSTAR have revealed that some parts of the spectrum of ULXP NGC~300 ULX-1 vary in response to the neutron star pulsation, and the variability can be categorized into two distinct phases. Since one was around the pulsation peak and the other was the rest, we named them the faint phase and bright phase, respectively. By applying the C3PO method developed by \citet{Noda2014} to ULXP for the first time, we successfully extracted two pairs of stable and variable component spectra that form the X-ray continuum of NGC~300 ULX-1 from each phase. Each spectral group, namely the stable, variable, and original spectra, was successfully reproduced with a combination of a disk blackbody (+Gaussian) and a Comptonization continuum with an extra high energy power law tail or wide absorption line. In the following paragraphs, we discuss what kind of physical origin may explain these observational results.

\subsection{Interpretation of the Stable Component in the Faint Phase \label{sec:interp_of_stable_comp}}
In this section, we give a possible implication to the origin of the convex non-pulsating spectral component extracted from the faint phase. According to theoretical studies on super-critically accreting objects, the entire accretion flow, from mass donating star to accretor, can be roughly separated into three regions in terms of distance from the center. One is the most outer region of the accretion flow where the radiative cooling is efficient enough to form an optically thick and geometrically thin standard accretion disk. In this region, the disk emits a multi-color disk blackbody spectrum with a radial surface temperature dependence of $\propto r^{-0.75}$ \citep{Shakura1973}. 

The closer to the center the stronger the emission becomes, and at a certain radius, the radiation pressure eventually overwhelms the self-gravitational pull of the disk. This forms the second flow region, in which the accretion disk starts to puff up and deviate from the ordinary standard disk regime. Within this radius, the accretion flow is now advection-dominated, in which a significant fraction of generated photons inside the disk are engulfed instead of being radiated from the disk surface. Although the region also emits a multi-temperature blackbody as the outer ``standard disk'' region, the radial temperature dependence is expected to be flatter as briefly described in section \ref{sec:spec_fit}, namely $\propto r^{-0.5}$ (e.g., \citealt{Watarai2000, Watarai2001}). In addition, a cool-optically-thick wind is expected to be launched from this region due to the intense radiation pressure (e.g., \citealt{Ohsuga2003, Kawashima2012}). The edge of this outflow can form a photosphere that emits an additional optically thick emission with a characteristic temperature. 

If the central object is a strongly magnetized neutron star, its magnetic pressure eventually overcomes the gas pressure of the flow at a certain point closer to the central object. Hence within this radius, the magnetic force restricts the accreting matters to move only along the magnetic field. This is the third characteristic region in the accretion flow, and it is called an accretion column (e.g., \citealt{Basko1975, Basko1976}). As the neutron star rotates, the entire column precesses along the magnetic field, accounting for the pulsating emission component. 

Since the stable component is unrelated to the pulsation, the emission clearly originates from some regions outside the accretion column, where the dipole magnetic field of the neutron star is too weak to capture the infalling matters. Therefore, it should be emission from either the outer or inner region of the accretion disk. According to the spectral analysis in section \ref{sec:fit_faint}, two types of thermal emission models, \texttt{diskbb} and \texttt{diskpbb}, gave acceptable fits. Since the single-temperature blackbody model failed to reproduce the spectra, we consider it unlikely that a photosphere of the outflow is the origin of the stable component. Although allowing the radial temperature profile to vary (using \texttt{diskpbb}) did not improve the fit and we could not statistically distinguish whether the data favor the disk to be in the standard or advection-dominated regime, we hereafter assume the former as a working hypothesis and proceed to further discussion. 

Since \texttt{diskbb} is an approximation of the standard accretion disk, its normalization gives an apparent inner-disk radius. The realistic radius $R_{\rm in}$ can be calculated by using the model normalization parameter $N$ as
\begin{equation}
    R_{\rm in}=\left(\frac{ND_{10}^2\xi^2\kappa^4}{\cos\theta_{\rm i}}\right)^{1/2}
    \label{eq:norm}
\end{equation}
(e.g., \citealt{Makishima2000}). Here, $D_{10}$, $\xi$, $\kappa$, and $\theta_{\rm i}$ represent the distance to the object in units of 10~kpc, a correction factor, color-hardening factor, and the inclination angle of the disk, respectively. 

The factor $\xi$ corrects for the difference in the inner-most edge boundary condition between \texttt{diskbb} and the standard accretion disk. The value is known to be $\xi=0.412$ \citep{Kubota1998} for a realistic standard accretion disk model that takes general relativity effects around a black hole into account. Due to the presence of the interrupting magnetic field, it is rather complex and challenging to estimate $\xi$ for the disk around an accreting neutron star. Some theoretical works suggest that the accretion disk may resemble the $\xi=0.412$ case, still, the value strongly depends on the details of the accretion flow (e.g., \citealt{Nixon2021}). Since $\xi$ for accreting neutron stars is thus unclear, here we assume the most popular value, $\xi=0.412$ \citep{Kubota1998}, as other studies on accreting neutron star low mass X-ray binaries (e.g., \citealt{Sakurai2014})

The color-hardening factor $\kappa$ is a ratio of the color temperature to the effective temperature. The value is known to be in between $\kappa=1.5\--2.0$ \citep{Shimura1995} at the sub-Eddington regime and recent numerical study suggests that the factor has a weak dependency on the mass accretion rate up to Eddington (e.g., \citealt{Davis2019}). Although the total mass accretion rate is in the a super-Eddington regime for a neutron star, here we assume that the local accretion mass density is still in the sub-Eddington level for the accretion flow at the radius where the disk is formed. Hence, we employ the most commonly used value of $\kappa=1.7$ \citep{Shimura1995}.

If we assume $D$ to be the distance to NGC~300 (1.9~Mpc; \citealt{Gieren2005}), then the present results in Table \ref{tab:fit_result} and Equation \ref{eq:norm} give a radius of $R_{\rm in}=720^{+220}_{-120}/\sqrt{\cos\theta_{\rm i}}$~km. Although the inclination angle of this system is still unknown, considering the detection of pulsation and the fact that no evidence of eclipse is present, we may assume that the system is nearly face-on to the observer ($\theta_{\rm i}\sim 0$). As discussed in section \ref{sec:interp_of_stable_comp}, $R_{\rm in}$ can imply the radius where the accretion flow shifts its behavior from the standard accretion disk to an accretion column; the magnetospheric radius $R_{\rm M}$.

In addition to the inner-disk radius, we can calculate the mass accretion rate at the radius by utilizing the theory of standard accretion disk \citep{Shakura1973} as
\begin{equation}
    \dot{M}=\frac{2R_{\rm in}}{GM}L_{\rm disk},
    \label{eq:mdot}
\end{equation}
where $G$, $M$, and $L_{\rm disk}$ are the gravitational constant, central object mass, and disk luminosity, respectively. Assuming a typical neutron star mass of $M=1.4M_{\rm \odot}$ and substituting values for $L_{\rm disk}$ and $R_{\rm in}$ from the present result, equation \ref{eq:mdot} gives a mass accretion rate of $\dot{M}\sim 1.3\times10^{20}$~g~sec$^{-1}$, which is $\sim50$ times the Eddington rate of $1.4M_{\rm \odot}$ neutron star ($\sim2.8\times 10^{18}$~g~sec$^{-1}$). In the following sections, we derive several physical values from these $R_{\rm in}$ and $\dot{M}$ and give a plausible explanation for the observed characteristics of NGC~300~ULX-1.

\subsubsection{Comparison With the Photon-trapping and Spherization Radii} \label{sec:r_trap}
In a super-critical accretion, two characteristic radii define where the accretion flow starts to deviate from the ordinary standard disk regime. One is a photon-trapping radius at which photons generated inside the flow start failing to escape from the surface of the accretion disk due to the strong advection derived from the high mass accretion rate. It is defined as the radius where the photon diffusion time scale becomes equivalent to the advection time scale (e.g., \citealt{Kato2008, Ohsuga2003}). The other is called the spherization radius, where the radiation pressure inside the flow overcomes the self-gravitational pull of the disk. According to several numerical calculations (e.g., \citealt{Shakura1973, Poutanen2007}, we can approximately obtain the latter as $R_{\rm sp}\sim 3\dot{m}R_{\rm S}$ where $R_{\rm S}$ and $\dot{m}$ are the Schwarzschild radius and the mass-accretion-rate ratio over the Eddington rate (i.e., $\dot{M}/\dot{M}_{\rm edd}=GM\dot{M}/6R_{\rm S}L_{\rm edd}$), respectively. Some studies also show that $R_{\rm sp}$ becomes nearly equivalent to the photon-trapping radius (e.g., \citealt{Poutanen2007}.)

Under the present accretion rate ($\dot{M}\sim 1.3\times10^{20}$~g~sec$^{-1}$), $R_{\rm sp}\sim R_{\rm trap}$ becomes $\sim 600$~km. Thus, the radii are smaller than $R_{\rm in}$, which indicates that the dipole magnetic field of the neutron star is bounding the accretion flow before it reaches the critical radius, from which the disk ``puffs up'' due to the radiation pressure. This may explain why the spectrum of NGC~300~ULX-1 exhibits fewer emission lines than sources that are considered to be forming a large-scale-height disk at their center, such as Swift J0243.6+6124 \citep{Bykov2022}.

A recent NuSTAR observation on a Galactic ULXP, Swift J0243.6+6124, has revealed that the source exhibits a significant Fe K$\alpha$ emission line in its spectrum at luminosity above the Eddington regime ($10.1\times10^{38}$~erg~sec$^{-1}$; \citealt{Bykov2022}). Since the Fe line component was reproduced with a reflection model that weakly varies over the pulse period, the authors proposed that the neutron star is embedded in a ``well'' formed by the inner edge of an inflated super-Eddington accretion disk. The central X-ray emission sweeps the inner wall as the neutron star rotates and generate the varying Fe fluorescence line. In contrast to Swift J0243.6+6124, we have found only a hint of the He-like Fe K$\alpha$ emission in the spectrum of NGC~300~ULX-1, and this can be due to an absence of such a geometrically-thick disk. Since we expect the strong dipole magnetic field to collimate the wall-illuminating emission toward the magnetic pole to some extent, the solid angle of the irradiating surface may decrease drastically if the disk is geometrically thin. In fact, the reflection fraction of Swift J0243.6+6124 decreased simultaneously with the X-ray luminosity, as the disk shifts its state toward the geometrically thin sub-Eddinton regime \citep{Bykov2022}. Hence, the relatively large $R_{\rm in}$ obtained from assumptions made in section \ref{sec:interp_of_stable_comp} seems convincing in terms of explaining the characteristics of its featureless spectrum.

\subsubsection{The Co-rotation Radius and the Observed X-ray Luminosity} \label{sec:r_co}
Since the magnetic field also rotates as the neutron star spins, we can define a radius where its rotation speed becomes equivalent to the Kepler velocity at that distance from the center. It is called a Co-rotation radius that can be derived by solving a balance between the spin period of neutron star $P$ and the Kepler motion period at radius $r$ as  
\begin{equation}
    R_{\rm c} = \left(\frac{GMP^2}{4\pi^2}\right)^{1/3}.
    \label{eq:r_c}
\end{equation}
If the system has a larger magnetospheric radius than its co-rotation radius, then the centrifugal force halts the accretion flow, and the entire system becomes X-ray dim. This phenomenon is called the propeller effect. Since ULX-1 is in an ultra-luminous phase, apparently, the effect is not taking place in this system. Therefore, the system must satisfy $R_{\rm M} < R_{\rm c}$ (in this case, $R_{\rm in} < R_{\rm c}$).

Under the current pulse period of ULX-1, $P\sim31$~sec, Equation \ref{eq:r_c} gives us a co-rotation radius of $R_{\rm c}=1.8\times 10^4$~km, which is significantly larger than $R_{\rm in}=720$~km. Hence, ULX-1 is obviously not suffering from the propeller effect, and it is consistent with its $>10^{39}$~erg~sec$^{-1}$ luminosity. Although the magnetospheric radius may vary depending on the mass accretion rate, the neutron star must spin up to a period of $P=0.32$~sec or shorter to achieve $R_{\rm c}<R_{\rm M}$ in this condition. According to a recent observation, the source has kept spinning up to $P\sim17$~sec and accreting matters at a similar rate as this observation \citep{Vasilopoulos2019}.

\subsubsection{An Estimation of the Magnetic Torque and Calculation of the Expected Spin-up Rate}
The matter infalling through the accretion disk can transfer its angular momentum to the central neutron star by applying torque onto the star through the magnetic ``arm'' that couples one to another. This forces the neutron star to spin up/down in time, and NGC~300 ULX-1 was, in fact, spinning up with a rate of $\dot{P}=5.56\times10^{-7}$~sec~sec$^{-1}$ within the present observation \citep{Carpano2018}. The torque applied to the neutron star with a moment of inertia $I$ and a spin angular momentum $\omega$ can be written as
\begin{equation}
    I\dot{\omega} = -2\pi I \frac{\dot{P}}{P^2}=\dot{M}\sqrt{GMR_{\rm M}}n(\omega_{\rm fast})
\label{eq:Iomega}
\end{equation}
 (e.g., \citealt{Ghosh1979b}, \citealt{Parfrey2016} \citealt{Vasilopoulos2018}) where $\dot{P}$ is the spin-up/down rate and $n(\omega_{\rm fast})$ is a function of dimensionless variable $\omega_{\rm fast}=(R_{\rm M}/R_{\rm C})^{3/2}$, which is known as the fastness parameter. For a slow rotator like NGC~300 ULX-1, namely $\omega_{\rm fast}\ll 1$, $n(\omega_{\rm fast})$ yields $\sim7/6$ \citep{Wang1995}. 

If we assume a typical neutron star ($M=1.4M_{\rm \odot}$ and a radius of $\sim10$~km) and a moment of inertia estimated from a certain equation of states and recent observational results (e.g., $I=1.6\times10^{38}$~m$^{2}$~kg; \citealt{Silva2021}), equation \ref{eq:Iomega} can be rewritten in terms of $\dot{P}$ as
 \begin{equation}
     \dot{P} = - 2.18\times10^{-8}\times P_{30}^2\dot{M}_{\rm edd}\sqrt{R_{720}}~{\rm sec~ sec^{-1}},
 \label{eq:pdot}
 \end{equation}
where $P_{30}$, $\dot{M}_{\rm edd}$, and $R_{720}$ are scaled parameters defined as $P_{30}=P/(30~{\rm sec})$, $\dot{M}_{\rm edd}=\dot{M}/(1.8\times10^{18}~{\rm g~sec^{-1})}$, and $R_{720}=R_{\rm M}/(7.20\times10^7~{\rm cm})$. Substituting values obtained from the present analysis, Equation \ref{eq:pdot} gives a spin-up rate of $\dot{P}=-1.9^{+1.4}_{-6.1}\times10^{-6}$~sec~sec$^{-1}$. Considering the mass distribution of the neutron star and the error obtained by \citet{Silva2021}, we here assumed that a $\sim30\%$ systematic error is present in $M$ and $I$. The derived value is consistent with the actually observed value $-5.56\times10^{-7}$ sec~sec$^{-1}$ within the error.

\subsubsection{Comparison of the Energy Budget within the Magnetosphere and the Observed X-ray Luminosity}
Let us test whether the observed X-ray luminosity is consistent with the expected value derived from the obtained mass accretion rate. As the intense magnetic field truncates the accretion disk at $R_{\rm M}$, the total energy budget for the emission from the inner-precessing flow should be equivalent to (or less than) the gravitational potential energy released between $R_{\rm M}$ and the neutron star surface. If we assume a free-fall accretion and the mass accretion rate to be constant at all radii, the observed luminosity is 
\begin{equation}
    L_{\rm obs}=\frac{GM\dot{M}_{\rm in}}{\gamma}\left(\frac{1}{R_{\rm NS}} - \frac{1}{R_{\rm M}}\right)
    \label{eq:lacc}
\end{equation}
where $R_{\rm NS}$, $\dot{M}_{\rm in}$, and $\gamma$ are the neutron star radius, the mass accretion rate at the inner-edge of the truncated disk, and the beaming factor, respectively. Since the dipole magnetic filed may collimate the emission pattern within $R_{\rm M}$ for some extent, a calculation assuming an isotropic radiation can amplify the apparent luminosity. We took this effect into account by scaling the value with a dimension-less factor $\gamma$ ($0< \gamma \le 1$).

Substituting the values derived from the present observation ($\dot{M}_{\rm in}=1.3\times10^{20}$~g~sec$^{-1}$, $R_{\rm M}=7.20\times10^{7}$~cm) and assuming a typical neutron star ($M=2.8\times10^{33}$~g, $R_{\rm NS}=10^{6}$~cm), equation \ref{eq:lacc} gives a luminosity of $(2.6/\gamma)\times10^{40}$~erg~sec$^{-1}$. Whereas the unabsorbed bolometric luminosity of the observed pulsating component is $1.1\times 10^{40}$~erg~sec$^{-1}$. We assumed the distance to the ULXP to be the same as that to NGC~300 (1.9 Mpc; \citealt{Gieren2005}) and an isotropic radiation. 
Although the true beaming factor is unknown, we may assume $\gamma$ to be close to unity because the pulse profile is rather sinusoidal (Figure \ref{fig:pulse_profile}) and an observation showed that the intensity of He II emission line in this system is consistent with that assuming emission with a minimal beaming effect \citep{Binder2018}.
Accordingly, the observed luminosity is roughly consistent with that expected from the accretion rate derived from an assumption that the origin of the stable component emission is an accretion disk.

\subsection{Magnetic Field Estimation}
Utilizing the obtained observational values, we estimate the strength of the dipole magnetic field of ULX-1 by following similar discussion as \citet{Walton2018a}. 
The magnetic field of a mass-accreting neutron star with a certain mass accretion rate $\dot{M}$ can be expressed as
\begin{equation}
    B=\left(\frac{R_{\rm M}}{2.6\times 10^6~{\rm cm}}\right)^{7/4}\dot{M}^{1/2}
    \label{eq:mag}
\end{equation}
\citep{Ghosh1979a, Lai2014, Furst2017}. Since $R_{\rm in}$ is now observable and $\dot{M}$ can be calculated via equation \ref{eq:mdot} utilizing observed $L_{\rm disk}$, we are able to derive the magnetic field $B$ by assuming a typical neutron star mass of $M=1.4M_{\rm \odot}$ from equation \ref{eq:mag}. 

\begin{figure}
    \centering
    \includegraphics[width=\columnwidth, bb=0 0 461 346]{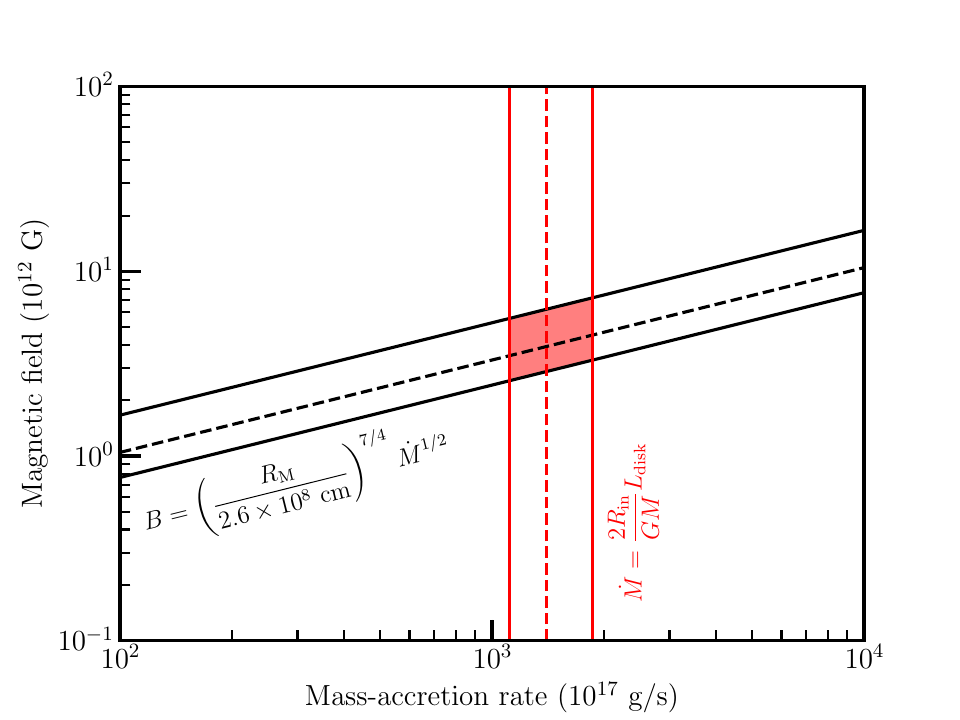}
    \caption{Estimated mass accretion rate and the magnetic field of ULX-1. Dashed lines are the best estimates, which are derived from equations \ref{eq:mag}, \ref{eq:mdot} and the best-fit values of the spectral analysis. Solid lines indicate the width of 90\% confidence level. The best estimate region is highlighted as the red-hatched region.}
    \label{fig:mdot_vs_B}
\end{figure}
In Figure \ref{fig:mdot_vs_B}, we present the relation between the mass accretion rate and the magnetic field derived from equations \ref{eq:mag} (black) and \ref{eq:mdot} (red). Each curve is drawn using the result obtained in the spectral analysis. According to this figure, our estimation of the magnetic field strength of ULX-1 lies in $2\--7 \times 10^{12}$~G (from the lowest to the highest tip of the red-hatched region). 

The estimated value is consistent with those made by other methods. For example, \citet{Vasilopoulos2018} estimated the strength of field as $5\times10^{12}$~G by utilizing the observed spin period evolution and a theory based on induced torque to neutron stars from accreting matters. \citet{Walton2018b} have found a possible absorption line feature that may be interpreted as a cyclotron resonance scattering feature. Assuming the feature originates from electron scattering, the authors concluded that the estimated magnetic field is $\sim10^{12}$~G. 

\subsection{Possible Picture of the Pulsating Accretion Flow}
Finally, let us discuss the possible structure of the pulsating accretion flow in NGC~300~ULX-1 from the variability we saw in the present analysis. As described in section \ref{sec:c3po} and section \ref{sec:spec_fit}, we have revealed that the variability of the pulsating component of this ULXP can be divided into at least two phases, the faint phase ($0.0 < \phi < 0.2$ and $0.5 < \phi < 1.0$) and the bright phase ($0.2\le\phi\le0.5$). 

In the faint phase, the pulsating component exhibited a hard continuum ranging from 0.5~keV to 25~keV with a rollover at $\sim7$~keV. Although we could roughly explain the spectrum with models that emit photons in a wide energy band with a characteristic cutoff temperature (multi-color disk blackbody, cutoff power law, and Comptonization), it required an extra component to explain the extending continuum at $> 10$~keV. The result is consistent with the previous studies of ULXs including the same ULXP (e.g., \citealt{Carpano2018, Walton2018a, Koliopanos2019}), and we employed an extra model to resolve this discrepancy. Simply adding a power-law model could not solve the residual at $> 10$~keV because the model extends to infinity in both energy directions and causes a conflict with the stable component that dominates the flux in the lower energy band. Thus, we concluded that the data favor a model that sharply drops off at the lower energy (e.g., Rayleigh-Jeans of the blackbody), and a model consisting of Comptonization plus its power-law scattering fits the data the best so far.

As discussed in the previous section, strongly magnetized neutron stars can form a magnetosphere, which forces the accreting matter to fall along its magnetic field. The magnetosphere creates a cylindrical-shaped accretion flow at a close region to the magnetic pole so-called accretion column (e.g., \citealt{Basko1975, Basko1976}). Within this column, free-falling matters are shock heated at a particular height from the stellar surface and emit high-energy X-ray photons. If the accretion rate gets close to the Eddington limit, these generated photons begin to escape from the sidewalls of the column rather than the direction of the magnetic field, creating an emission pattern perpendicular to the magnetic field called a ``fan beam'' \citep{Basko1976}. Some of these fan-beam photons can irradiate an optically thick region trapped around the boundary of the magnetosphere. The incident photons will experience photoelectric absorption and multiple Compton scatterings within this region, and some eventually escape the system as reprocessed thermal emission (e.g., \citealt{Mushtukov2017}). Other fractions of the fan beam photons may instead irradiate the neutron star surface close to the magnetic pole and be reflected as a secondary emission pattern called a ``polar beam'' (e.g., \citealt{Trumper2013, Poutanen2013}). In this case, photons are re-emitted parallel to the magnetic field with a spectrum harder than the original fan beam. Thus, the accretion flows around highly magnetized mass accreting neutron stars can form multiple emission regions, and several observations on other ULXPs (e.g., \citealt{Koliopanos2018} for SMC X-3, and \citealt{Bykov2022} for Swift J0243.6+6124) have supported such a complicated emission geometry. Although the sources were not in the ultra-luminous state, the resent X-ray polarization observations on accreting neutron star binaries (\citealt{Tsygankov2022} for Cen X-3, and \citealt{Marshall2022} for 4U 1626-67) have also hinted the presence of the multi-zone pulsed emission.

Some numerical studies suggest (e.g., \citealt{Mushtukov2017, Mushtukov2019}) that at well above the Eddington rate, the reprocessing region can extend further down to the neutron star to obscure the central hard X-ray emitting region. As almost the entire region within the magnetosphere has shifted to the reprocessing area, the accretion flow at this rate is no longer regarded as a column but rather as a ``curtain'' (for example, see Figure 1 in \citealt{Mushtukov2017}). It shields and reprocesses most of the hard X-ray photons from the central fan beam into thermal black-body-like emission. Since this reprocessing optically thick curtain has a particular temperature gradient along the magnetic field, one can approximate its overall spectrum with the summation of black bodies from respective temperature regions. A numerical study estimates the lowest and highest temperature of the black bodies to be around sub keV and a few keV for a neutron star accreting at $5\times10^{39}$~erg~sec$^{-1}$ with a magnetic field of $10^{13}$~G \citep{Mushtukov2017}. The estimated value is roughly consistent with the present observational result. The variable component spectrum exhibited breaks at both ends in energy, and we successfully explained the lower one with the Rayleigh-Jeans of 0.16 keV blackbody. Although the Comptonization model, not the multi-color blackbody as expected in the theoretical study, gave the best fit to the data, we consider that the model just happened to be the one to approximate the actual temperature gradient of this accretion flow. The smooth single-peaked pulse profile (Figure \ref{fig:pulse_fraction}) suggests that the emission from the other side of the pole is not reaching the observer, which also supports the idea that the reprocessing region covers most of the magnetosphere.  

According to several theoretical studies, including numerical simulations (e.g., \citealt{Ohsuga2005, Ohsuga2007, Ohsuga2011, Kawashima2012, Jiang2014, Sadowski2014}), an extreme accretion rate as ULXPs may derive the surrounding disk to launch optically thick and cool outflows. Since this creates a tall funnel-like structure around the central source, it can easily interfere with the pulsating emission and change its spectral shape by Compton down scattering or absorbing photons as the entire accretion curtain precesses around the spin axis. \citet{Kosec2018} reported the presence of blue-shifted absorption lines in the same data set with a $\sim3\sigma$ significance. The detection is evidence of highly-ionized matters outflowing with $\sim20\%$ of the speed of light. However, the spectrum of the pulsating component does not change its shape as the intensity varies within each phase interval (see the CCPs in section \ref{sec:c3po}). The behavior indicates that the emission appearance of the individual accretion curtain regions, which accounts for the intensity change in each phase interval, is somehow uniform and does not depend on the rotation phase; suggesting only the solid angle of the emission region to the observer changes as the curtain rotates with the neutron star. Furthermore, \citep{Kosec2018} estimated the column density of the outflow to be $1.2^{+1.9}_{-0.6}\times10^{23}$~cm$^{-2}$, which is relatively transparent to the Thomson scattering ($<10^{24}$~cm$^{-2}$). According to several numerical simulations of super-critical accretions, such Thomson-thin gas can leak out from the accretion curtain \citep{Abolmasov2022} and be accelerated up to $10\--40\%$ of the speed of light via intense radiation pressure and magnetic reconnections near the magnetic pole \citep{Takahashi2017}. 
Considering the observational and simulation results above, we propose that, instead of the disk, the intense radiation of the rotating accretion curtain is blowing the thin matter leaked out near the neutron star away. Since the strong magnetic field is likely truncating the disk before it derives the geometrically/optically thick super-critical winds (see section \ref{sec:r_trap}), the central curtain emission is visible without being interfered with by the matters within the disk. 

Although the theoretical studies thus qualitatively describe most of the spectral behavior of the pulsating component below 10~keV, we still do not have a good explanation for the tail component above that. Due to the limited signal-to-noise ratio above 25~keV, it is rather challenging to determine whether the feature above 10~keV originates from an absorption line argued by \citet{Walton2018a} or an extending power-law continuum. We expect a future hard X-ray mission to resolve this problem.

\begin{figure}
    \centering
    \includegraphics[width=\columnwidth]{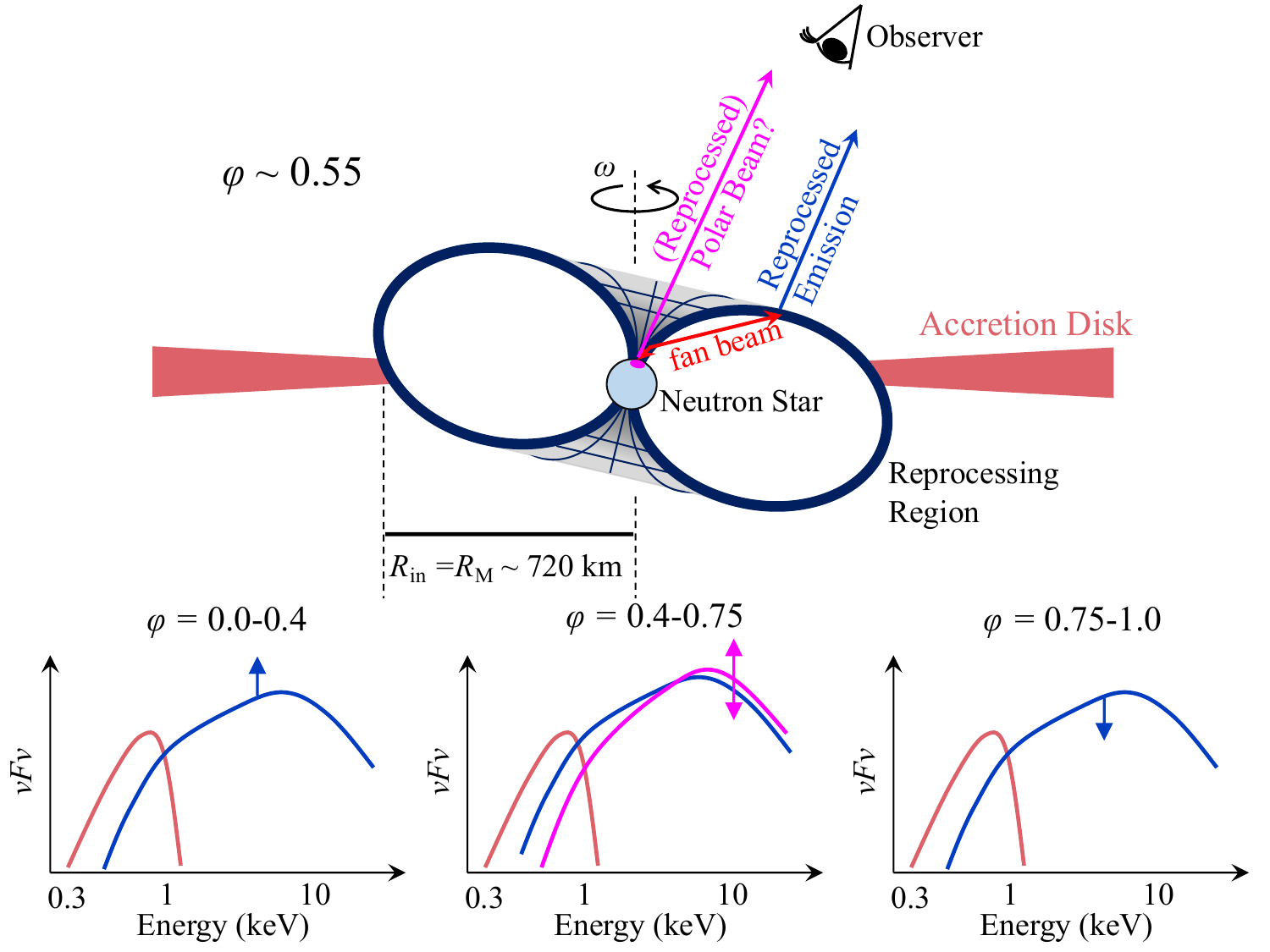}
    \caption{Schematic pictures of the possible accretion geometry (top) and spectral variability (bottom) of NGC~300~ULX-1. }
    \label{fig:sche_ngc300}
\end{figure}

As the source reaches the bright phase (or the pulse peak), the component that used to be variable in the faint phase (blue solid line in Figure \ref{fig:sche_ngc300}) halts to increase its intensity, and an alternative variable component (magenta solid line in Figure \ref{fig:sche_ngc300}) with a similar but slightly harder continuum emerges. It strongly indicates that the accretion flow accounting for the pulsation consists of at least two representative emission regions. Furthermore, considering that the harder component is observable only in the bright phase ($\sim 30\%$ of the pulsation period), the emission can be slightly collimated toward the magnetic pole axis. Such a multi-region structure has never been reported in the rotating accretion component of ULXPs.

In the top half of Figure \ref{fig:sche_ngc300}, we present a schematic cross-section drawing of a possible accretion geometry that explains our observational result. As described above, an optically thick curtain is likely formed along the magnetic field due to the super-Eddington accretion. Since the magnetic field confines the accretion flow into a small region near the magnetic pole, the entire curtain creates a funnel structure, which obscures most of the neutron star and the central hard X-ray (fan beam) emitting region from the observer. Such curtains around magnetized neutron stars are also suggested by theoretical studies including numerical simulations (e.g., \citealt{Takahashi2017, Mushtukov2019, Abarca2021, Inoue2023}). The entire curtain exhibits a reprocessed multi-color blackbody spectrum shown in blue lines. The area near the stellar surface (the magenta region) is observable only when the opening cone faces the direction toward the observer as the flow precesses. Thus, the hard variable component shown in the magenta line emerges in a limited phase ($0.2 \le \phi \le 0.5$), whereas the blue reprocessed component is present in the spectrum at all times.

Since this emerging continuum is harder than that in the faint phase, it can be interpreted as an emission from the polar beam. This is because the polar beam is a scattered-off fan beam component, in which the lower energy part of the original emission experiences a photoelectric absorption at the neutron star surface. The emission of the polar beam typically extends up to $>10$~keV (e.g., \citealt{Koliopanos2019, Trumper2013}), whereas that of NGC~300~ULX-1 rolls over at lower energy as $\sim7$~keV. Since the spectral shape is relatively similar to the variable component in the faint phase, the polar beam can be also reprocessed by the optically thick curtain, and we might have observed the remnant of its original hard spectrum. Recent two-dimensional simulation indicates \citep{Kawashima2020} that the super-critical accretion flow onto a highly magnetized neutron star can form a complex structure inside its opening funnel. Also, pulse shape and pulse fraction depends on the the structure of the funnel and observer's viewing angle \citep{Inoue2020}. Although the expected spectrum from such a complex structure is yet to be available, the present result might support such a simulation.

Finally, we discuss how the present accretion scenario is related to a longer time-scale spectral evolution. According to the recent SWIFT/XRT and NICER monitoring campaign on ULX-1, the source has continuously decreased its X-ray flux by an order of magnitude since this observation in 2016 (e.g., \citealt{Ray2019, Vasilopoulos2019, Ng2022}). Although the cause of the dimming is still unclear, \citet{Vasilopoulos2019} suggested that obscuration by radiation pressure-dominated disk and its outflow can explain the phenomenon rather than a decrease in intrinsic mass accretion rate. This is because the source kept spinning up at the same rate as that in the brightest era, meaning the central neutron star continuously gained an equivalent amount of accreting matters, i.e., spin-up torque, in the dimming phase. \citet{Ng2022} also supported this scenario by reporting spectral softening and occasional detections of possible partially-covered disk emissions during this X-ray flux decrease. As we calculated in section \ref{sec:r_trap}, the estimated $R_{\rm M}$ of ULX-1 is comparable to or marginally larger than $R_{\rm sp}$ (or $R_{\rm trap}$) in this particular observation. Therefore, a slight increase in mass accretion rate may result in a formation of a geometrically thick radiation pressure-dominated disk region that launches thick outflows. Given that the intrinsic mass accretion rate has been comparable to or possibly higher than the present data, we suggest that a radiation-pressure-dominated disk  has formed during the dimming phase due to a further decrease of $R_{\rm M}$ (or an increase of $R_{\rm sp}$) from the 2016 observation and such a thick disk has obscured the central neutron star region as it precesses along the line of sight due to a physical mechanism; such as Lense-Thirring precession (e.g., \citealt{Middleton2017}) as \citet{Vasilopoulos2019} and \citet{Ng2022} proposed.

\section{Summary and Conclusions}
We reanalyzed X-ray data sets of NGC~300~ULX-1 taken with XMM-Newton and NuSTAR on 2016/12/16. In addition to the classical phase-resolved spectral analysis, we newly employed a method that has been developed in the AGN studies called the C3PO method. As previous studies have reported, the pulse profile of NGC~300~ULX-1 is relatively sinusoidal and peaked at one particular phase throughout the observed energy band, suggesting a single-zone emission region from its appearance. However, the C3PO method has revealed that the pulsating emission varies differently within $\pm15\%$ of the pulsation peak. The result suggests that, instead of being a single hot zone, the pulsating accretion flow consists of at least two representative emission regions. Accordingly, we divided the entire data into these two phase intervals and performed further C3PO analysis procedures for each, separately.

For each phase interval, the C3PO method provided an extra pair of spectra that represent the actual shape of the components consisting of the overall spectrum of NGC~300~ULX-1. One coincides with the pulsation and the other does not. Thanks to this extra spectral information that was not available in the previous studies, we have successfully put more stringent restrictions on the spectral modelings and resolved several model degeneracies that have been reported in the same data set. The stable component is explained with a geometrically thin standard disk model with $0.25\pm0.03$~keV peak temperature and a $720^{+220}_{-120}$~km inner radius. As for the variable component, the best-fit model is a combination of Comptonization of a $0.14\pm0.04$ keV blackbody and its extra up-scattered power law. It disfavors some models that were successful in the previous studies, especially those extending infinitely to the lower energy band. Furthermore, its continuum is found to be slightly hard in the brighter phase.

The obtained disk parameters gave a mass accretion rate of $1.3\times10^{20}$~g~sec$^{-1}$, which is $\sim50$ times the Eddington rate of a $1.4M_{\rm \odot}$ mass neutron star. Using this accretion rate and assuming the disk-inner radius to be equivalent to the magnetospheric radius, we estimated the spin-up rate, the X-ray luminosity of the rotating flow, the spherization radius, and the dipole magnetic field strength of this system. All of the values are consistent with the observed values, numerical simulations, and estimations made by other independent methods. Considering the results, we have proposed that the system consists of a geometrically thin accretion disk truncated by a $2\--7\times 10^{12}$~G magnetic field and an inner-precessing accretion flow exhibiting a hard continuum with $1.1\times10^{40}$~erg~sec$^{-1}$ luminosity. The latter flow is likely forming a funnel-like geometry with two representative temperature regions, and its inner-hot part is observed only when the opening cone points toward the observer.


\begin{acknowledgments}
The authors would like to thank all of the members of XMM-Newton and NuSTAR teams for their devotion to instrumental calibration and spacecraft operation. This research has been supported by JSPS KAKENHI grant numbers JP19K21054, JP19K21884, JP20H01941, and JP20H01947, in part by JSPS Grant-in-Aid for Scientific Research (A) JP21H04488 and Multidisciplinary Cooperative Research Program in CCS, University of Tsukuba. This work was also supported by MEXT as “Program for Promoting Researches on the Supercomputer Fugaku” (Toward a unified view of the universe: from large-scale structures to planets, JPMXP1020200109), and by Joint Institute for Computational Fundamental Science (JICFuS).
\end{acknowledgments}

%

\vspace{5mm}
\facilities{XMM, NuSTAR}


\software{NUMPY \citep{Harris2020}, ASTROPY \citep{Astropy2018}, VEUSZ (https://veusz.github.io/), MATPLOTLIB \citep{Hunter2007}, ROOT (\url{https://root.cern/}), HEASOFT (\url{https://heasarc.gsfc.nasa.gov/docs/software/heasoft/}), and SAS (\url{https://www.cosmos.esa.int/web/xmm-newton/download-and-install-sas})}




\bibliography{bondi}{}
\bibliographystyle{aasjournal}

\appendix
\restartappendixnumbering
\section{Count-Count Plot of FPM-B}
\begin{figure}[htb!]
    \centering
    \includegraphics[width=\columnwidth, bb=0 276 907 842]{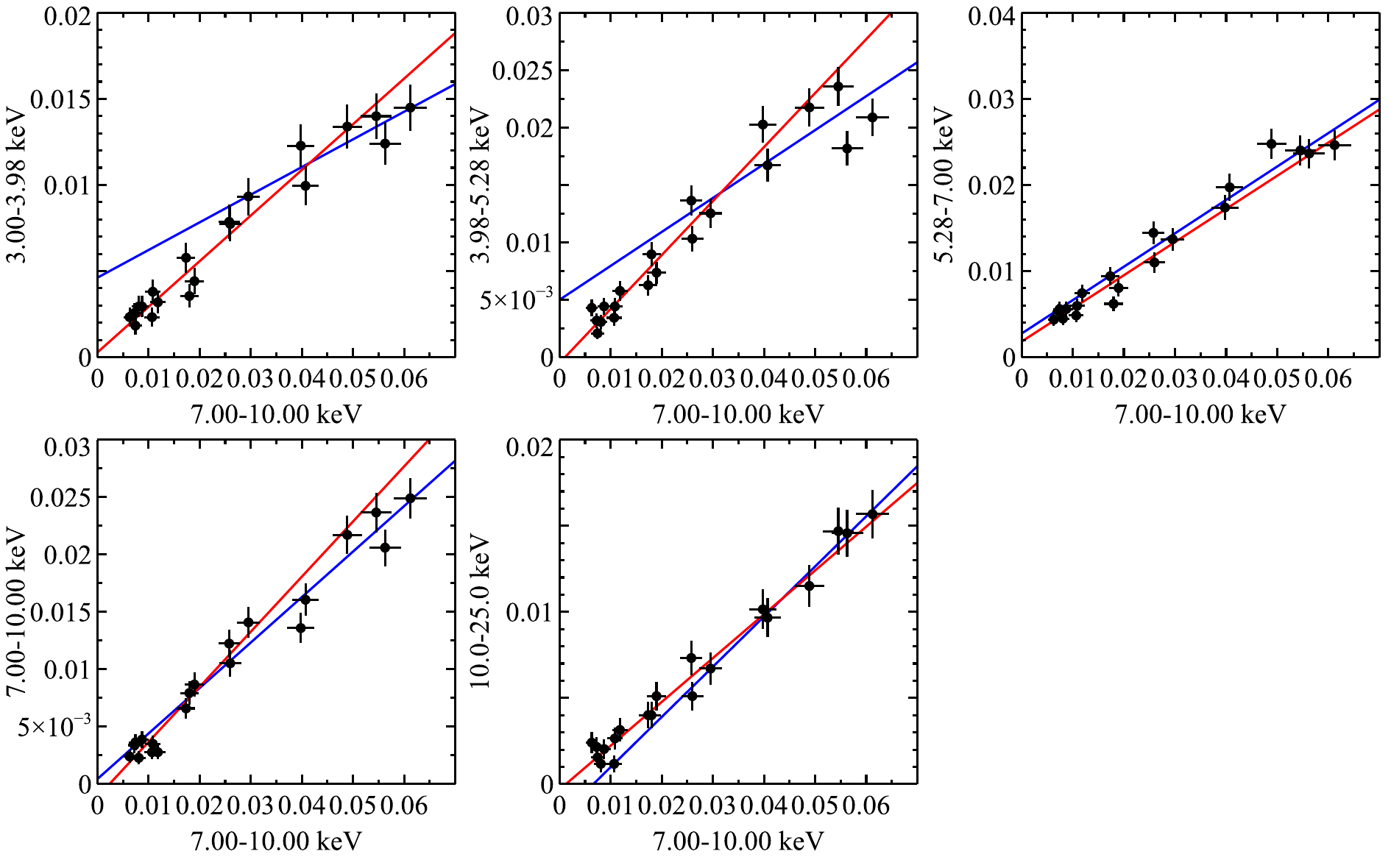}
    \caption{Same as figure \ref{fig:ccplot_fpma}, but the data points indicate the count rate of NuSTAR FPM-B.}
    \label{fig:ccplot_fpmb}
\end{figure}


\end{document}